\documentclass[fleqn,usenatbib]{mnras}

\usepackage[T1]{fontenc}

\DeclareRobustCommand{\VAN}[3]{#2}
\let\VANthebibliography\thebibliography
\def\thebibliography{\DeclareRobustCommand{\VAN}[3]{##3}\VANthebibliography}

\usepackage{latexsym}
\usepackage{graphicx}
\usepackage{amssymb}
\usepackage{longtable}
\usepackage{epsf}
\usepackage{amsmath}
\usepackage{textcomp}
\usepackage{gensymb}

\usepackage{hyperref}
\hypersetup{colorlinks=true,linkcolor=blue,citecolor=blue,filecolor=blue,urlcolor=blue}
\usepackage{cleveref}
\usepackage{bm}		

\usepackage{natbib}
\usepackage{comment}

\newcommand{\msun}{{\rm M}_{\odot}}

\newcommand{\beq}{\begin{eqnarray}}
\newcommand{\eeq}{\end{eqnarray}}
\newcommand{\ben}{\begin{enumerate}}
\newcommand{\een}{\end{enumerate}}
\newcommand{\bit}{\begin{itemize}}
\newcommand{\eit}{\end{itemize}}

\newcommand{\dd}{\rm d}

\newcommand{\dmqdt}[1]{{\rm d}#1/{\rm d}t}
\newcommand{\dmqdtfrac}[1]{\frac{{\rm d}#1}{{\rm d}t}}
\newcommand{\dmqdtpfrac}[1]{\frac{{\rm d}#1}{{\rm d}t'}}

\newcommand{\Msun}{{\rm M}_{\odot}}
\newcommand{\Mpeak}{M_{\rm peak}}
\newcommand{\Mhalo}{M_{\rm halo}}
\newcommand{\Mh}{M_{\rm h}}

\newcommand{\Mgas}{M_{\rm g}}
\newcommand{\Mstar}{M_{\star}}
\newcommand{\Mstart}{M_{\star}(t)}

\newcommand{\mzero}{M_{0}}

\newcommand{\tacc}{t_{\rm acc}}

\newcommand{\mh}{m_{\rm h}}
\newcommand{\rvir}{R_{\rm vir}}

\newcommand{\terf}{\mathcal{T}_{\rm erf}}

\newcommand{\dtime}{{\rm d}t}

\newcommand{\mseff}{\epsilon_{\rm ms}}

\newcommand{\taudep}{\tau_{\rm dep}}
\newcommand{\taucons}{\tau_{\rm cons}}
\newcommand{\sfr}{{\rm d}\Mstar/{\rm d}t}
\newcommand{\sfrms}{{\rm d}\Mstar^{\rm ms}/{\rm d}t}
\newcommand{\sfrt}{{\rm d}\Mstart/{\rm d}t}
\newcommand{\sfrtfrac}{\frac{{\rm d}\Mstart}{{\rm d}t}}
\newcommand{\sfrtfracms}{\frac{{\rm d}M_{\star}^{\rm ms}(t)}{{\rm d}t}}

\newcommand{\fcons}{F_{\rm cons}}
\newcommand{\Mcrit}{M_{\rm crit}}
\newcommand{\mcrit}{m_{\rm crit}}
\newcommand{\Qfunc}{F_{\rm q}}
\newcommand{\qfunc}{f_{\rm q}}
\newcommand{\qdrop}{q_{\rm drop}}
\newcommand{\qrejuv}{q_{\rm rejuv}}
\newcommand{\qtime}{t_{\rm q}}

\newcommand{\qdt}{q_{\rm dt}}
\newcommand{\qfrac}{F_{\rm q}}
\newcommand{\rejuvfrac}{F_{\rm rejuv}}

\newcommand{\rhosfr}{\rho_{\rm SFR}}
\newcommand{\rhogas}{\rho_{\rm gas}}

\newcommand{\tauc}{\tau_{\rm c}}
\newcommand{\aearly}{\alpha_{\rm early}}
\newcommand{\alate}{\alpha_{\rm late}}

\newcommand{\phalf}{p^{\rm halo}_{\rm 50\%}}

\newcommand{\tform}{t_{\rm form}}

\usepackage[usenames,dvipsnames,table]{xcolor} 
\definecolor{hpurple}{HTML}{7E16DF}

\definecolor{horange}{HTML}{FFA500}

\usepackage{xspace}
\newcommand{\tng}{IllustrisTNG\xspace}
\newcommand{\um}{UniverseMachine\xspace}
\newcommand{\dstarcode}{{\tt diffstar}\xspace}
\newcommand{\dstar}{{Diffstar}\xspace}

\newcommand{\DstarPop}{{DiffstarPop}\xspace}
\newcommand{\dmah}{{Diffmah}\xspace}

\title[Diffstar: Parametric Galaxy Assembly History]{Diffstar: A Fully Parametric Physical Model for Galaxy Assembly History}

\author[Alarcon et al.]{
Alex Alarcon$^{1}$\thanks{E-mail:alexalarcongonzalez@gmail.com},
Andrew P. Hearin$^{1}$,
Matthew R. Becker$^1$,
Jon\'{a}s Chaves-Montero$^{1,2}$
\\
$^1$HEP Division, Argonne National Laboratory, 9700 South Cass Avenue, Lemont, IL 60439, USA\\
$^2$Donostia International Physics Centre, Paseo Manuel de Lardizabal 4, 20018 Donostia-San Sebastian, Spain
}

\date{Accepted XXX. Received YYY; in original form ZZZ}

\pubyear{2022}

\begin{document} 
\label{firstpage} \pagerange{\pageref{firstpage}--\pageref{lastpage}} \maketitle

\begin{abstract}
We present \dstar, a smooth parametric model for the in-situ star formation history (SFH) of galaxies. \dstar is distinct from conventional SFH models that are used to interpret the spectral energy distribution (SED) of an observed galaxy, because our model is parametrized directly in terms of basic features of galaxy formation physics. The \dstar model assumes that star formation is fueled by the accretion of gas into the dark matter halo of the galaxy, and at the foundation of \dstar is a parametric model for halo mass assembly, Diffmah. We include parametrized ingredients for the fraction of accreted gas that is eventually transformed into stars, $\mseff,$ and for the timescale over which this transformation occurs, $\taucons;$ some galaxies in \dstar experience a quenching event at time $\qtime,$ and may subsequently experience rejuvenated star formation. We fit the SFHs of galaxies predicted by the \tng (TNG) and \um (UM) simulations with the \dstar parameterization, and show that our model is sufficiently flexible to describe the average stellar mass histories of galaxies in both simulations with an accuracy of $\sim0.1$ dex across most of cosmic time. We use \dstar to compare TNG to UM in common physical terms, finding that: (i) star formation in UM is less efficient and burstier relative to TNG; (ii) galaxies in UM have longer gas consumption timescales, $\taucons$, relative to TNG; (iii) rejuvenated star formation is ubiquitous in UM, whereas quenched TNG galaxies rarely experience sustained rejuvenation; and (iv) in both simulations, the distributions of $\mseff$, $\taucons$, and $\qtime$ share a common characteristic dependence upon halo mass, and present significant correlations with halo assembly history. We conclude the paper with a discussion of how \dstar can be used in future applications to fit the SEDs of individual observed galaxies, as well as in forward-modeling applications that populate cosmological simulations with synthetic galaxies.
\end{abstract}




\section{Introduction}
\label{sec:intro}

One of the core goals of extragalactic astronomy is to understand the relationship between the fundamental physical parameters of a galaxy and its observed spectral energy distribution (SED). Stellar population synthesis (SPS) is the prevailing framework that enables theoretical predictions for the SED of a galaxy \citep{conroy13_sps_review}, and the star formation history of a galaxy (SFH) is one of the fundamental physical properties that determines its SED. There are several distinct approaches that are commonly used to model SFH. In traditional parametric approaches, some simple functional form is assumed for the shape of ${\rm SFH}(t),$ and the parameters of this functional form are programmatically varied in the SPS analysis. Some examples of typical parametric models are functions that are exponentially declining \citep{Schmidt1959}, delayed exponential \citep{Sandage1986_sfhs}, lognormal \citep{Gladders2013, diemer_etal17}, and double power-laws \citep{Behroozi2013, Ciesla2017, Carnall2018}. There are numerous alternatives to such simple functional forms. For example, it is increasingly common to use a piecewise-defined model\footnote{Note that it is common practice to refer to these models as ``non-parametric". As pointed out in \citet{leja_etal19}, this is a misnomer: in these models, star formation history is deterministically specified by the values of ${\rm SFH}(t)$ at the control points. In this paper, we will refer to these as  ``piecewise-defined models", and reserve the term ``non-parametric" for models in which there truly does not exist an exact parametric description of ${\rm SFH}(t)$, such as a hydrodynamical simulation or semi-analytical model.} that characterizes ${\rm SFH}(t)$ by interpolating between a set of control points in time \citep{cid_fernandes_etal05,Ocvirk2006,chauke_etal18,tojeiro_etal07,leja_etal19,iyer_etal19}. There are also a range of alternatives such as stochastically correlated models \citep{caplar_tachella19, tacchella2020_StochasticModellingstarformation}, models formulated in terms of a set of basis functions \citep{iyer_gawiser_2017,Sparre2015, Matthee_Schaye2019, Chen2020_empiricalSFH}, and entirely non-parametric approaches such as drawing SFHs directly from a simulation of galaxy formation \citep{finlator_etal07,Pacifici2012,Pacifici2015}.

When conducting an SPS analysis of the SED or photometry of a galaxy, the choice one makes for the SFH model has significant consequences for the inference of a galaxy's physical properties. First, it is important for the SFH model to have sufficient flexibility such that the galaxy properties of interest are not biased by the underlying assumptions. On the other hand, the model should not be so flexible that the constraining power of the observational data becomes unacceptably degraded. This ubiquitous trade-off between bias and variance has been discussed extensively in the literature on SFH models. For example, in \citet{simha_etal14} the authors introduced a 4-parameter SFH model that they tested against star formation histories taken from a hydrodynamical simulation, and demonstrated that their model only slightly inflates statistical errors relative to one-dimensional models, but with the benefit of a major reduction in systematic biases. In closely related work analyzing the Simba simulation \citep{dave_etal19_simba}, in \citet{lower_etal20} it was found that piecewise-defined models are able to infer star formation rates with much smaller biases relative to simple parametric forms such as a delayed-$\tau,$ again with only a modest inflation of the parameter posteriors.

A significant challenge to observational studies of star formation history is that inferring the detailed shape of galaxy SFH is a fundamentally under-conditioned problem. Even when considering high-resolution measurements of galaxy spectra ($R\sim10,000$) with signal-to-noise ratios as large as 100, only $\sim8$ distinct episodes of star formation can be discerned \citep{Ocvirk2006}. These limitations are even more severe for less detailed measurements: multiple studies have now shown that Bayesian analyses of photometry are commonly prior-dominated \citep{Carnall18,leja_etal19,lower_etal20}. This highlights the potential danger of over-interpreting the observations based on a model whose complexity is unwarranted by the available data, and so in all SPS analyses of star formation history, careful consideration of the assumed prior is critical.

Motivated by these considerations, in this paper we introduce a new parametric approach to modeling galaxy SFH, \dstar. Our  functional form assumes a physically-motivated relationship between the assembly of the underlying dark matter halo, the efficiency of star formation along the main sequence, the consumption timescale of freshly accreted gas, and the possibility of a quenching event. We use simulated star formation histories from the \um \citep{behroozi_etal19} and \tng \citep{Pillepich2018a, Springel2018} to validate the flexibility of our parameterization, and show that \dstar supplies a compact description of these simulations that enables a simple, physics-based comparison of their predicted SFHs.

This paper is organized as follows. In \S\ref{sec:sims}, we describe the simulated datasets used throughout the paper. We give a pedagogical overview of the \dstar model in \S\ref{sec:model_formulation}, and in \S\ref{sec:model_perform} we assess the performance of our model's ability to capture the SFHs in the \um and \tng simulations. In \S\ref{sec:interpretaish}, we study the statistical trends and scaling relations exhibited by \um and \tng, and we use our model as the basis of a physical comparison between these two simulations. We discuss our findings and future applications of \dstar in \S\ref{sec:discussion}, and conclude in \S\ref{sec:conclusions} with a summary of our principal results.

\section{Simulations}
\label{sec:sims}

In order to validate that the \dstar model for star formation history (SFH) is sufficiently flexible, we used simulated SFHs taken from publicly available datasets based on \tng \citep[TNG,][]{Nelson2018b} and \um \citep[UM,][]{behroozi_etal19}. We now describe these two synthetic datasets in turn.

\tng is a suite of cosmological hydrodynamical simulations that incorporates a wide variety of baryonic feedback processes, including radiative gas cooling, star formation, galactic winds, and AGN feedback \citep{Weinberger2017, Pillepich2018a}. We use publicly available data from the largest hydrodynamical simulation of the suite, TNG300-1. The TNG300-1 simulation was carried out using the moving-mesh code {\sc Arepo} \citep{Springel2010} to solve for the evolution of $2500^3$ gas tracers together with the same number of dark matter particles in a simulation box of 302.6 Mpc on a side, under a cosmology very similar to \citet{planck14b}. For TNG300-1, the corresponding mass resolution is $5.9\times10^7\Msun$ and $1.1\times10^7\Msun$ for dark matter and gas, respectively. Halos and subhalos in \tng were identified with the {\tt SUBFIND} algorithm \citep{springel_etal2001_subfind}, and the merger trees we use were constructed with {\tt SUBLINK} \citep{rodriguez_gomez_etal15_sublink}. Publicly available galaxy properties are tabulated at 100 snapshots. The snapshot spacing ranges from $36\, {\rm Myr}$ to $193\, {\rm Myr}$, with a median spacing of  $153\, {\rm Myr}$.

\um is an empirical model of galaxy star formation across redshift; at each simulated snapshot, the UM model maps a value of SFR onto every subhalo, and the SFH of the simulated galaxies are determined in a post-processing analysis of the merger trees. For the synthetic SFHs used in this paper, we use the best-fit model of \um run on the Bolshoi Planck simulation \citep[BPL,][]{klypin_etal11, klypin_etal16}. The BPL simulation was carried out using the ART code \citep{kravtsov_etal97_art} by evolving $2048^3$ dark-matter particles of mass $m_{\rm p}=2.2\times10^{8}\Msun$ on a simulation box of $368.7\,{\rm Mpc}$ on a side, under cosmological parameters closely matching \citet{planck14b}. Merger trees were identified with Rockstar and ConsistentTrees \citep{behroozi_etal13_rockstar, behroozi_etal13_consistent_trees, rodriguez_puebla_etal16} prior to running the \um code. The 178 UM snapshots have a spacing that ranges from $15\, {\rm Myr}$ to $124\, {\rm Myr}$, with a median spacing of  $80\, {\rm Myr}$.

All results in the paper pertain to the assembly histories of present-day host halos (i.e., {\tt upid=-1} for Rockstar, and the ``main halo'' for {\tt SUBFIND}). Our choice to focus on the in-situ assembly history of central galaxies is an important simplifying feature of our analysis, particularly regarding in the interpretation of galaxy quenching in lower-mass halos; we will separately study the phenomenon of satellite quenching and merging in follow-up work. 

Throughout the paper, including the present section, values of halo mass, stellar mass and distance are quoted assuming the Hubble parameter used by each simulation ($h_{\rm BPL}=0.678$ and $h_{\rm TNG}=0.6774$).

For notational convenience, throughout the paper, all references to the $\log$ function will be understood to refer to base-10 logarithms, without exception. We will use the variable $m_h$ to denote $\log M_h\ [M_\odot]$, and we will use the notation ${\rm sSFR}\equiv\log\dot{\rm M}_{\star}/\Mstar\ [{\rm yr^{-1}}].$

When writing $\Mstar$, stellar mass or stellar mass history (SMH) throughout the text, we refer to the total mass formed in stars, or cumulative star formation, without taking into account mass that has been lost due to the finite lifetimes of stars.

\section{Diffstar Model Formulation}
\label{sec:model_formulation}

In this section, we give a detailed description of the parametric formulation of the \dstar model. This includes both a formal definition of the model, as well as theoretical motivation for each ingredient; quantitative justification for each ingredient based on individual examples of simulated SFH appears in this section; we present additional justification for our formulation based on fits to large samples of simulated galaxies in the following section. We refer the reader to Appendix~\ref{appendix:model_def} for a concise summary of the \dstar parametrization.

In the modern picture of galaxy formation, stars form from gaseous baryonic matter that is gravitationally bound within dark matter halos, and this process is accompanied by a litany of feedback mechanisms that operate across a large dynamic range of scales in space and time \citep[see][for contemporary reviews]{mo_vdb_white_2010,somerville_dave_2015,Vogelsberger2020}. Through a diverse body of evidence ranging from hydrodynamical simulations \citep{Schaye15,Khandai2015,feldmann_etal16,Pillepich2018a}, semi-analytic models \citep{kauffmann_etal99,kang_etal05,Croton2006,benson_galacticus_2012}, and empirical techniques \citep{conroy_wechsler_2009,watson_hearin_age_matching3,behroozi_silk_2015,becker_smad_2015,moster_emerge1,behroozi_etal19}, it is now well established that star formation is a relatively inefficient process, and exhibits strong correlations with both the mass and the assembly history of the parent halo of the galaxy. Our goal with \dstar is to develop a parametric form that is flexible enough to capture the diversity of pathways by which galaxies and halos co-evolve, and that at the same reflects the underlying simplicity of the most fundamental aspects of the galaxy--halo connection.

In the basic physical picture of the \dstar model, we assume that baryonic matter becomes available for star formation at a rate that is closely related to the growth rate of the dark matter halo. In \S\ref{subsec:model_mah}, we review our model for halo mass assembly history, which is the same as the \dmah model presented in \citet{hearin_etal21_diffmah}, to which we refer the reader for additional details. In \S\ref{subsec:wot_gas_accretion}, we motivate and discuss our use of \dmah to approximate the rate at which {\it baryonic} mass becomes available for star formation.

Once gas falls into the dark matter halo, in \dstar we make the ansatz that only a fraction of the accreted material ever transforms into stars, and that this fraction depends only upon the instantaneous mass of the parent halo. We furthermore assume that the mass-dependence exhibits a characteristic shape, such that there is a critical mass where the conversion fraction peaks, and that at lower and higher halo masses the conversion fraction falls off monotonically. We describe this modeling ingredient in detail in \S\ref{subsec:wot_bce}.

In real galaxies, stars do not form instantaneously at the first moment that a parcel of gas falls inside the boundary of a dark matter halo, and so in \dstar we assume that there is a characteristic timescale, $\taucons,$ over which an accreted parcel of gas will gradually be transformed into stellar mass. We refer to $\taucons$ as the {\it gas consumption timescale}, and we discuss this aspect of our model in \S\ref{subsec:wot_depletion}. 

In the \dstar model, there is parametrized flexibility to capture the phenomenon that some galaxies experience a quenching event that results in a pronounced reduction in star formation. Furthermore, the flexibility of our model allows for the possibility that quenching is not permanent, and that some galaxies experience rejuvenated star formation after having been previously quenched. We discuss how \dstar treats these two phenomena in \S\ref{subsec:wot_quenching}. 

\subsection{Halo Mass Assembly}
\label{subsec:model_mah}
\begin{figure}
\includegraphics[width=8cm]{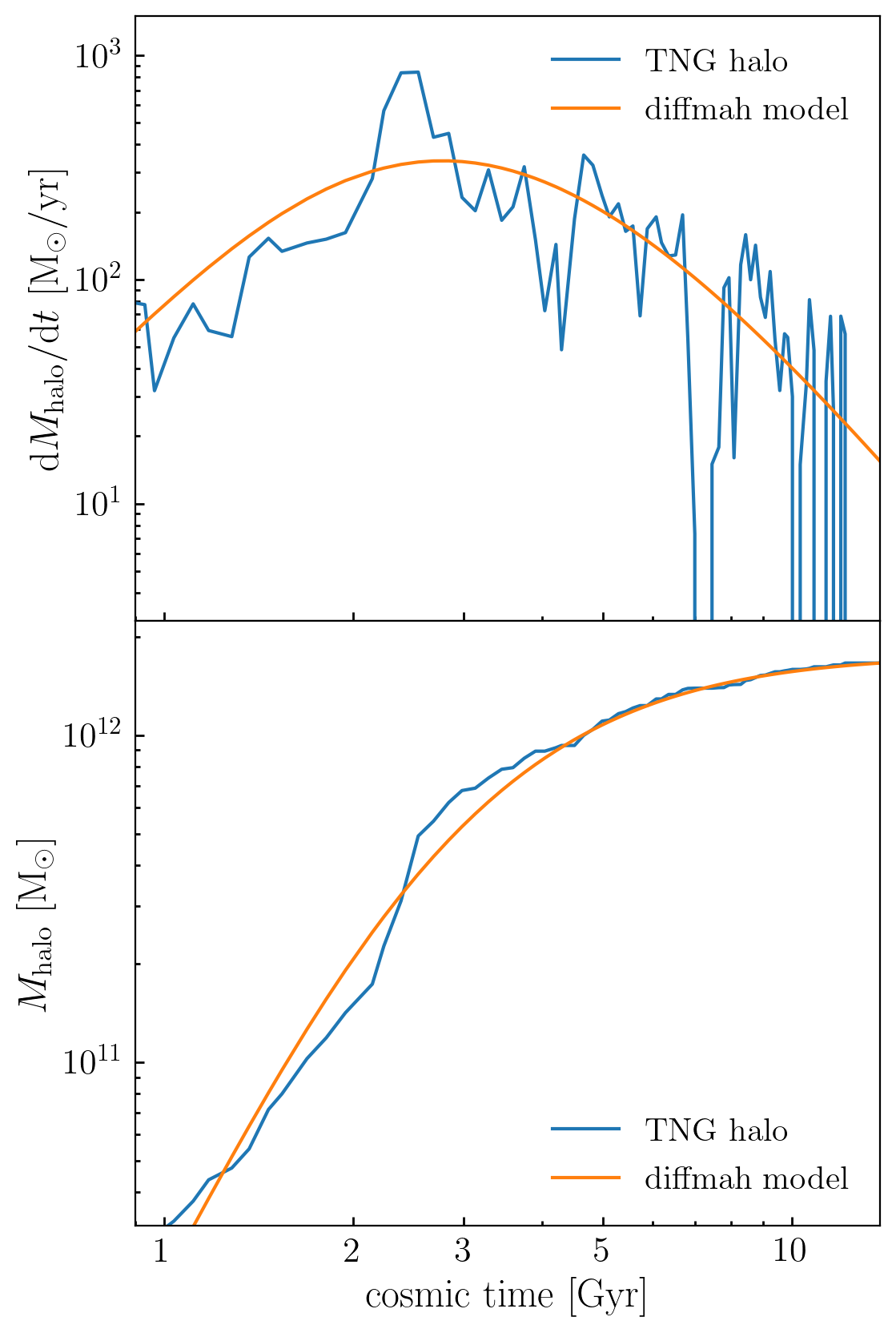}
\caption{{\bf Diffmah model of halo mass assembly}. In each panel, the blue curve shows the assembly history of the same halo in \tng, and the orange curve shows the approximate halo history supplied by \dmah. The top panel shows the mass accretion rate, and the bottom panel shows the cumulative mass.}
\label{fig:diffmah}
\end{figure}

At the foundation of the \dstar model is the mass assembly history (MAH) of the dark matter halo hosting the galaxy. We model the MAH using the \dmah parameterization presented in \citet{hearin_etal21_diffmah}, in which $\Mhalo(t)$ is defined to be a power-law function of cosmic time with a rolling index,
\beq
\label{eq:individual_mah}
\Mhalo(t) = \mzero(t/t_0)^{\alpha(t)},
\eeq
where $t_0$ is the present-day age of the universe, and $\mzero\equiv\Mhalo(t_0).$ We model the time-dependence of the power-law slope using a sigmoid function, $\mathcal{S}(x),$ defined as follows:
\beq
\label{eq:sigmoid}
\mathcal{S}(x\vert x_0,k,a, b) = a + \frac{b-a}{1 + \exp\{-k(x-x_0)\}}.
\eeq
The function $\mathcal{S}(x)$ is smoothly differentiable and monotonically increases from $a$ to $b$ with a characteristic transition at $x_0.$ Thus in Eq.~\ref{eq:individual_mah}, the behavior of $\alpha(t)$ is given by
\beq
\label{eq:diffmah_slope}
\alpha(t)= \aearly + \frac{\alate-\aearly}{1 + \exp\{-k_{h}(t-\tauc)\}}.
\eeq
The parameters $\aearly$ and $\alate$ define the asymptotic values of the power-law index at early and late times, respectively; $\tauc$ is the transition time between the early- and late-time indices, and $k_{h}=3.5\ {\rm Gyr^{-1}}$ defines the speed of the transition, and is held constant for all halos. 

The \dmah model describes $\Mhalo(t)$ to be a smooth, monotonically increasing function of time, and so this model is intended to approximate the history of the {\it peak} halo mass, ensuring $\dmqdt{\Mhalo}$ is everywhere non-negative. Hard-wired into the functional form of \dmah and its fitting procedure is the modern physical picture of dark matter halo growth: at early times, halos undergo a period of rapid growth during a ``fast-accretion regime", and mass accretion rates diminish considerably at later times as halos transition to the ``slow-accretion regime". Even though the \dmah model imposes this narrative onto the interpretation of simulated merger trees, the three free parameters of the model have sufficient flexibility to capture the diversity of MAHs of individual halos in either gravity-only or hydro simulations, with a typical accuracy of $\sim0.1$ dex across time \citep[see][for further details]{hearin_etal21_diffmah}. We provide further discussion below of how our choice to use \dmah as the basis of halo growth influences the \dstar formulation. Figure~\ref{fig:diffmah} shows an example \dmah fit to the mass assembly history of a dark matter halo in \tng.

\subsection{Gas Accretion Rate} 
\label{subsec:wot_gas_accretion}

As stated above, the \dstar model assumes that the rate at which baryonic mass becomes available for star formation in a galaxy is closely related to the growth rate of the parent dark matter halo. Although this is rather intuitive at a qualitative level, it is not clear a priori whether this assumption is suitable for the level of quantitative analysis that we intend to carry out. After all, dark matter appears to be dissipationless, and is only subject to gravitational forces, whereas gas is collisional, and so it can shock, mix, and dissipate energy via radiative cooling. And as shown in \citet{van_der_voort_etal11} and \citet{faucher_giguere_etal11}, stellar winds, outflows, and other baryon-specific processes have potential to significantly impact the accretion rates of gas versus dark matter, particularly in the inner regions of a halo, close to where the actual galaxy resides. Moreover, for purposes of predicting star formation rate, there is a questionable physical basis for the adoption of commonly-used boundaries of dark matter halos such as the virial radius, $\rvir,$ because the definition of $\rvir$ is tied to a reference background density that evolves with time, which in turn can lead to inferred growth of halo mass even if the physical density profile of the halo remains constant \citep{diemer_etal13,more_etal15_splashback_radius_is_physical}. 

However, numerous works carefully studying the physical nature of the accretion of gas into halos has revealed a strikingly close connection between the assembly history of baryonic and dark matter within individual halos. In \citet{dekel_etal13}, the authors used a suite of cosmological zoom-in simulations to identify broadly similar baryonic accretion rates at $\rvir$ and $\rvir/10.$ This finding was confirmed and strengthened in \citet{wetzel_nagai_2015}, who found that the physical accretion rate of baryons at {\it all} radii
within the halo roughly tracks the accretion rate across the halo boundary. It has also been shown in \citet{Mitchell2021} that the internal and ejected gas of halos in the EAGLE simulation approximately follows the cosmic baryon fraction, $f_{\rm b}.$ Motivated by these results, we make the following assumption in the \dstar model,
\beq
\label{eq:dmgasdt}
\dmqdtfrac{\Mgas} = f_{\rm b}\dmqdtfrac{\Mh},
\eeq
where $\dmqdt{\Mgas}$ is the accretion rate of baryonic material that is available for star formation, and $\dmqdt{\Mh}$ is the growth rate of total halo mass. 

In Eq.~\ref{eq:dmgasdt}, we use the \dmah model to approximate $\dd{\Mh}/\dd t.$ The fact that halo growth in \dmah is smooth  has important implications for the formulation and interpretation of \dstar, particularly regarding the sharp transient fluctuations in $\dmqdt{\Mh}$ that are a characteristic feature of numerical estimations of halo mass growth from simulated merger trees. Some of these fluctuations correspond to physical events such as major mergers that could impact the halo's resident galaxy \citep[see, e.g.,][]{wang_zentner_etal_2020}, but there are also quite significant timestep-to-timestep fluctuations for which the connection to the physics of galaxy formation is tenuous. Using simulated merger trees for $\dmqdt{\Mh}$ is tantamount to an at-face-value interpretation of each individual fluctuation in an N-body merger tree as corresponding to a true, physical fluctuation in the in-situ star formation rate of the galaxy. The direct use of simulated trees furthermore introduces an unwanted dependence of the model upon the resolution of the simulation, both in terms of the particle mass and the spacing of the snapshots in time. In \dstar, we use smooth approximations to $\dmqdt{\Mh}$ based on \dmah, thereby neglecting short-term fluctuations that appear in simulated merger trees; in \S\ref{sec:discussion}, we discuss a future extension of our model that will incorporate such fluctuations in a manner that is not tied to transient fluctuations in simulated merger trees.

\subsection{Baryon Conversion Efficiency} 
\label{subsec:wot_bce}

\begin{figure}
\includegraphics[width=8cm]{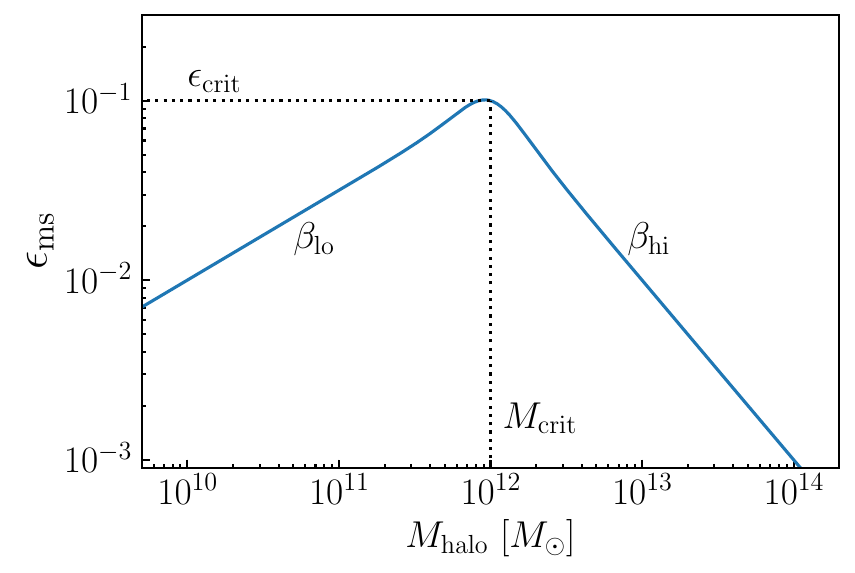}
\caption{{\bf Baryon conversion efficiency function}. The value of $\mseff$ controls the fraction of accreted gas that is transformed into stars: $\delta\Mstar\propto\mseff\times\delta\Mgas.$ The \dstar model assumes that $\mseff(\Mhalo)$ varies with halo mass according to a characteristic shape, presenting a peak in efficiency determined by some characteristic halo mass $M_{\rm crit},$ and falling off like a power law at lower and higher halo mass. The figure gives a visual illustration of the form of the parametric freedom given to the shape of  $\mseff(\Mhalo).$ See Eqs.~\ref{eq:mseff}-\ref{eq:diffstar_slope} and accompanying text for details.}
\label{fig:sfreff}
\end{figure}

\begin{figure*}
\centering
\includegraphics[width=0.9\textwidth]{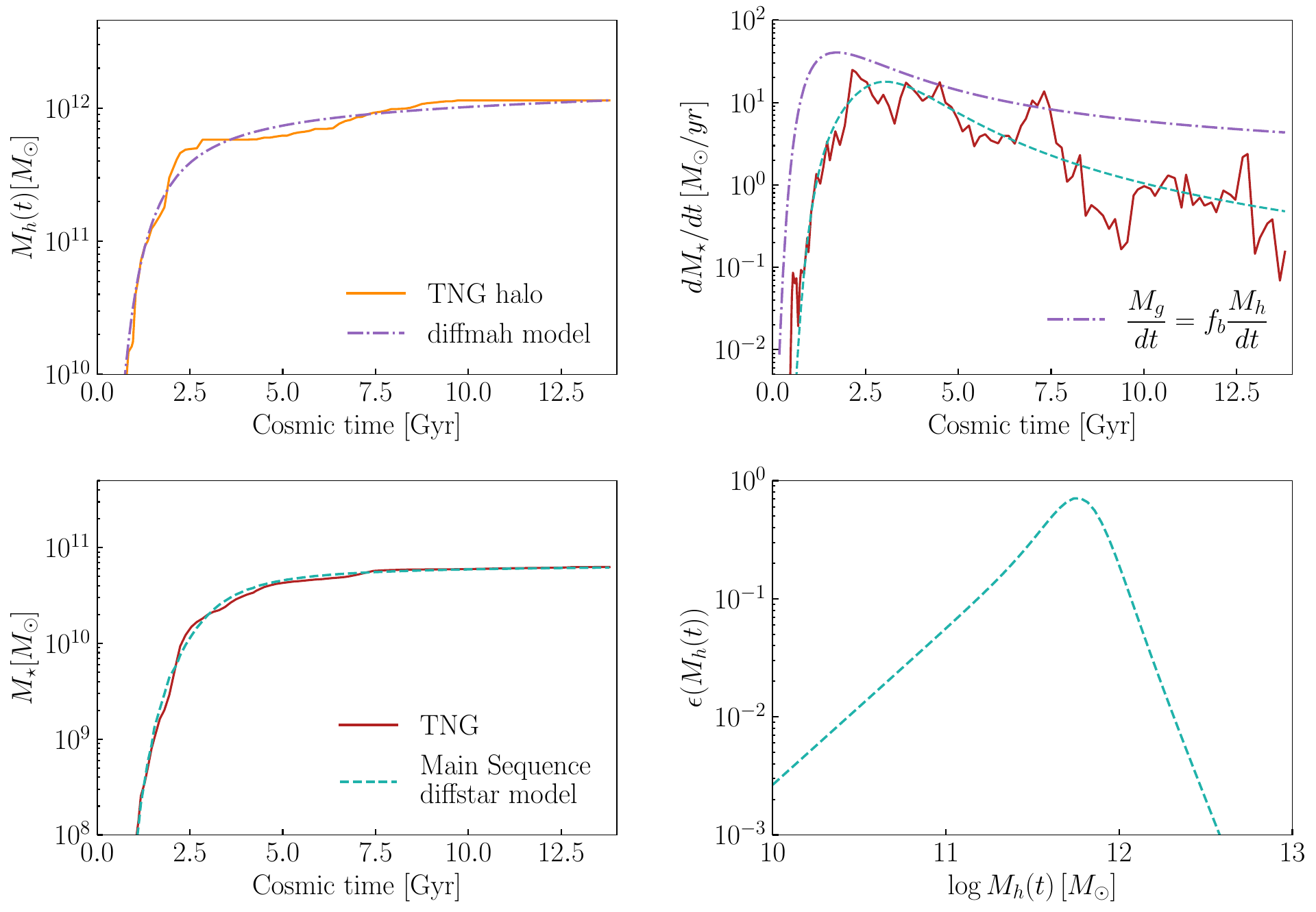}
\caption{{\bf Example fit to the assembly history of an individual galaxy/halo in TNG}. 
{\it Top left:} 
The mass assembly history of the simulated halo (solid line), 
and the approximation based on \dmah (dot-dashed line).
{\it Top right:} 
The star formation history of the simulated galaxy (solid line), the accretion rate of gas implied by the \dmah fit (dot-dashed), and the star formation history of the best-fitting \dstar model (dashed). 
{\it Bottom left:} 
The stellar mass history of the galaxy.
{\it Bottom right:} 
The baryon conversion efficiency of the best-fitting \dstar model.
}
\label{fig:wot_moster}
\end{figure*}

Once a parcel of gas falls inside the boundary of a dark matter halo, only a fraction of the accreted mass ever ultimately transforms into stars. For a small parcel of gas, $\delta\Mgas,$ that accretes at some time, $t,$ the portion of this mass that turns into stars at some later time, $t'>t,$ is controlled by the {\em baryon conversion efficiency}, $\mseff(\Mh(t')),$ which is defined by the following proportionality:
\beq
\label{eq:bce}
\delta\Mstar(t')\propto\mseff(\Mh(t'))\times\delta\Mgas(t).
\eeq

In formulating this problem as in Eq.~\ref{eq:bce}, we adopt a similar approach as in \citet{mutch_etal13}, and make the ansatz that $\mseff$ depends only upon the total mass of the parent halo at the moment that the gas is converted into stars, and that the form of $\mseff(\Mhalo)$ peaks at some critical mass, $M_{\rm crit},$ and falls off monotonically at lower and higher halo masses. This ansatz is motivated by a wide range of evidence. In lower mass halos, hydrodynamical simulations and semi-analytic models have shown that stellar winds from massive stars and supernovae can eject gas from the shallow potential of the halo, thereby reducing the total amount of baryonic material that is available to fuel the formation of stars \citep{nelson_etal19_tng50_outflows}; additionally, star formation can reheat cold gas, creating conditions that prevent further conversion of baryonic matter into stars \citep{benson_etal02,Hopkins2012}; other physical processes such as photoionization of the intergalactic medium are also thought to play an important role in inhibiting star formation in low-mass halos \citep{benson_etal02}. Meanwhile, higher-mass halos host massive black holes that can be very effective at preventing star formation, either by the heating of the surrounding gas and/or the ejection of the gas from the galaxy, or by the creation of kinetic bubbles that impart momentum to the surrounding gas \citep{Croton2006, Sijacki2007, Gabor2010, Weinberger2017, Fluetsch2019, Trussler2020}. 

Further evidence for our assumed shape of $\mseff(\Mhalo)$ comes from results based on empirical models. One of the basic findings of abundance matching studies is that the shape of the dark matter halo mass function together with the shape of the observed stellar mass function requires that $\Mstar/\Mhalo,$ the ratio of stellar mass to halo mass, has a peak near $\Mhalo\approx10^{12}\Msun,$ and falls off towards lower and higher halo mass \citep{Moster2010,Behroozi2010,moster_etal13,Behroozi2013a}. This general shape appears to be nearly redshift-independent across most of cosmic time \citep{behroozi_etal2013b}, which strongly suggests that the efficiency of star formation has a similar shape.

Motivated by these considerations, we model the halo mass dependence of the baryon conversion efficiency as
\beq
\label{eq:mseff}
\mseff(\Mh) = \epsilon_{\rm crit}\cdot(\Mh/\Mcrit)^{\beta(\Mh)}.
\eeq
Thus $\mseff(\Mh)$ behaves like a power law with a rolling index, $\beta(\Mh)$. The efficiency attains its critical value\footnote{Note that $\epsilon_{\rm crit}$ is close, but not quite equal to the peak value of $\mseff(\Mh),$ due to the functional form defined by Eq.~\ref{eq:mseff}.} of $\epsilon_{\rm crit}$ when the host halo mass equals $\Mcrit$. To model the $\Mhalo$-dependence of $\beta,$ we use the same functional form shown in Eq.~\ref{eq:sigmoid} for a sigmoid function:
\beq
\label{eq:diffstar_slope}
\beta(\mh)= \beta_{\rm lo} + \frac{\beta_{\rm hi}-\beta_{\rm lo}}{1 + \exp\{-k(\mh-\mcrit)\}},
\eeq
where as described in \S\ref{sec:sims}, for notational convenience we have written $\mh=\log\Mh,$ and $\mcrit=\log\Mcrit$, and $k=9$ is held constant. Figure~\ref{fig:sfreff} gives a visual illustration of $\mseff(\Mh)$. We note that our assumed form for $\mseff(\Mhalo)$ is similar to the one adopted in \citet{moster_emerge1}, and has four parameters\footnote{In practice, when fitting the SFHs of individual halos with the \dstar model, we hold fixed $\beta_{\rm hi}=-1$ after we include the possibility of a quenching event; see \S\ref{subsec:wot_quenching} and \S\ref{subsec:fitting_procedure}.} controlling the behavior of the $\Mhalo$-dependence: $\{\Mcrit,\,\epsilon_{\rm crit},\,\beta_{\rm lo},\,\beta_{\rm hi}\}$.

Figure~\ref{fig:wot_moster} shows an example fit of the \dstar main sequence model for a halo with Milky Way mass at $z=0$. The top left panel shows the mass assembly history of an \tng halo, with the best-fitting approximation from \dmah. In the top right panel, the solid red curve shows the star formation history of an example galaxy from \tng. The dot-dashed purple curve in the top right panel shows $\dd{\Mgas}/\dd t,$ which is modeled as in Eq.~\ref{eq:dmgasdt}; when we determine the halo mass accretion rate used to define $\dd{\Mgas}/\dd t,$ both in this figure and throughout the paper, we use the best-fitting \dmah approximation to the values of $\dd{\Mhalo}/\dd t$ taken from the simulated merger tree. The dashed cyan curve in the top right panel shows best-fitting \dstar approximation to the simulated SFH. In the particular fit shown with the dashed cyan curve in this panel, we assume that star formation history is simply given by $$\sfrt=\mseff(\Mh(t))\cdot\dd{\Mgas}(t)/\dd t,$$ so that the conversion of baryonic mass into stars happens instantaneously at the moment the gas falls inside the virial radius of the halo (see \S\ref{subsec:wot_depletion} below for how we relax this assumption in the full \dstar formulation). The bottom left panel compares the stellar mass history of the simulated galaxy to its best-fitting approximation, and the behavior of $\mseff(\Mh)$ of the best-fitting model is shown in the bottom right panel.

\subsection{Gas Consumption Timescale}
\label{subsec:wot_depletion}

\begin{figure}
\includegraphics[width=8cm]{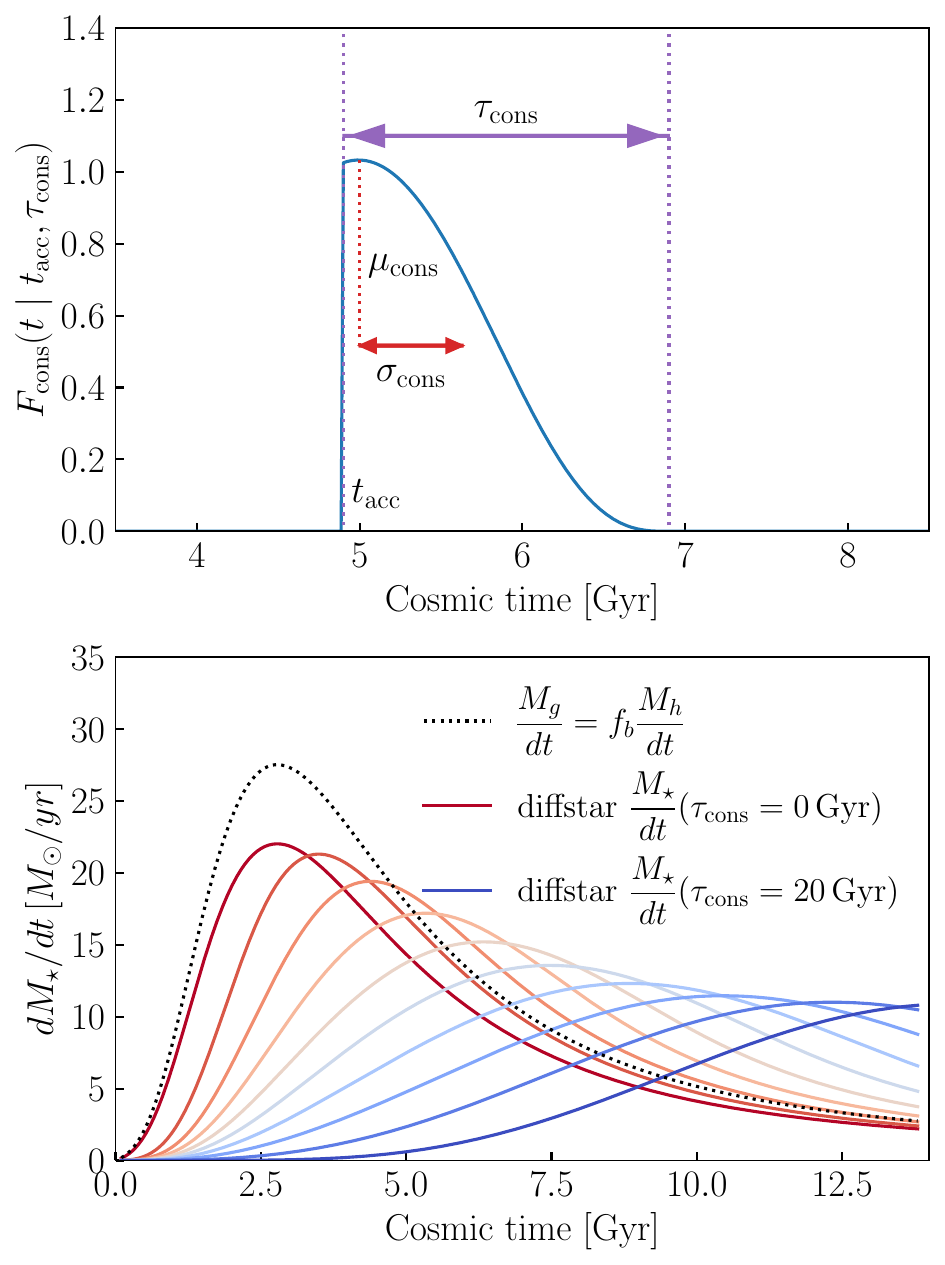}
\caption{{\bf Effect of the gas consumption function on SFH.}
{\it Top panel:} consumption function. For a parcel of gas accreted at time $\tacc,$ the behavior of $\fcons(t)$ distributes the  stars formed from the gas over the timescale, $\taucons.$ See Eqs.~\ref{eq:mssfr}-\ref{eq:fdep} and text for details.
{\it Bottom panel:} SFH of an early-forming halo with $\mseff=0.8,$ and different values of $\taucons.$ For larger $\taucons,$ the SFH peaks at later times.
}
\label{fig:depletion}
\end{figure}

\begin{figure}
\centering
\includegraphics[width=8cm]{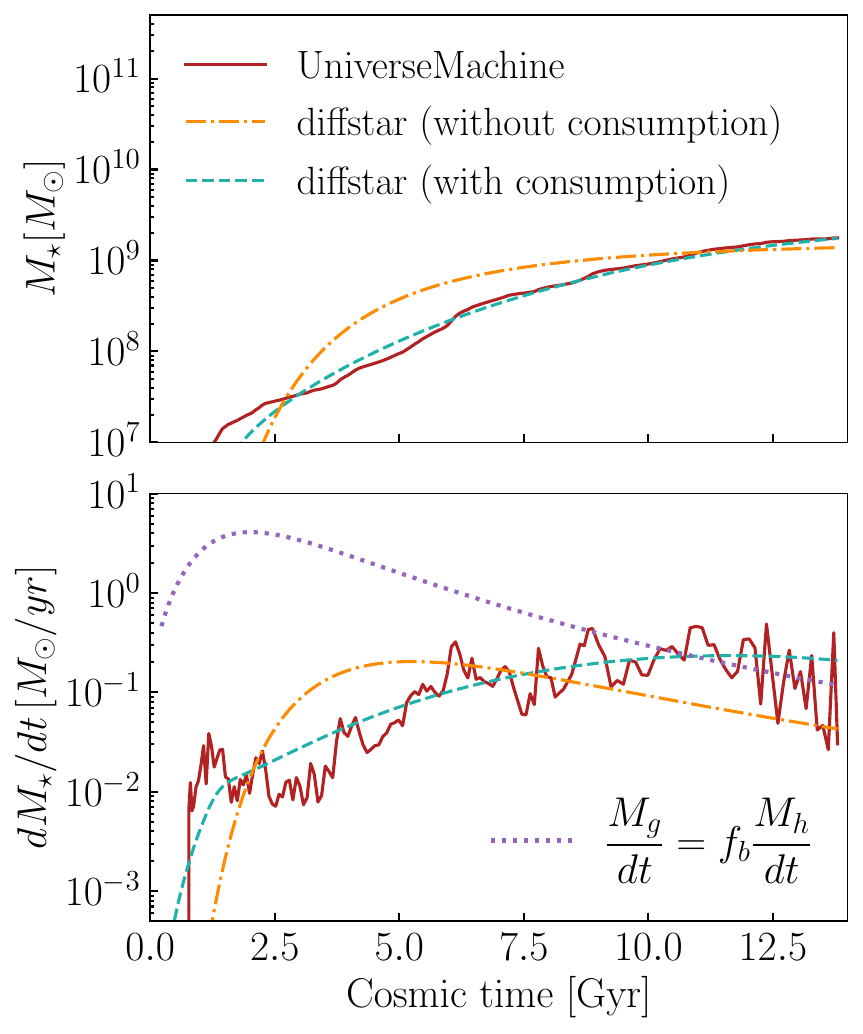}
\caption{{\bf Gas consumption timescale and SFH fits.} 
{\it Top panel:} Stellar mass history. {\it Bottom panel:} History of star formation rate (SFH).
The solid red lines show the assembly history of a galaxy in UM. The dot-dashed orange lines show an approximation to this history based on a version of the \dstar model in which $\taucons$ is held fixed to zero, so that accreted gas instantaneously transforms into stars according to some best-fitting efficiency. The dashed cyan lines show the best-fitting \dstar model in which $\taucons$ is permitted to vary as a free parameter, allowing for gas to gradually turn into stars over time. The dotted purple line in the bottom panel shows the gas accretion history, which is proportional to the mass accretion rate given by the best-fitting \dmah approximation. The gas accreted by this early-forming halo turns into stars gradually over several Gyr, reproducing the late-time peak in SFH.
}
\label{fig:wot_depletion}
\end{figure}

When a dark matter halo accretes a fresh parcel of gas from the field, there can be a considerable lag in time before this parcel cools down and forms the molecular clouds that fuel star formation. Indeed, there is considerable evidence from observations of the Milky Way and nearby spiral galaxies that this lag can be quite long, with timescales ranging $1\sim10\, {\rm Gyr}$ \citep{Kennicutt1989, Kennicutt1998, Bigiel2008, Leroy2008, Leroy2013, reyes2019, Diaz2020, Kennicutt2021}. Evidence for very long gas consumption timescales also comes from detailed analyses of high-resolution hydrodynamical simulations. In simulations of isolated disk galaxies, it was shown in \citet{Semenov2017, semenov_etal18} that gas cycles rapidly between star-forming and non-star-forming states, with only a small fraction of the gas being converted into stars in any one cycle, such that the many cycles are needed before the gas reservoir becomes depleted. It was furthermore found that gas in the interstellar medium (ISM) cycles between these states due to a combination of effects: gas compression/expansion when entering/exiting spiral arms; stellar/supernova feedback that disperses star-forming regions and generates large-scale ISM turbulence; shocks from expanding SNe bubbles that compress gas in the disk plane, thereby inducing new star-forming regions and subsequent SNe explosions; and gas ejected in fountain-like outflows that eventually falls back due to the gravitational pull of the disk. Long gas consumption timescales, with high variance from halo to halo, are thus a natural consequence of this physical picture.

In \dstar, we parametrize this phenomenon in terms of $\taucons,$ the timescale over which an accreted parcel of gas will be gradually transformed into stellar mass. Thus in our model, the star formation rate of a galaxy, $\sfrt,$ receives a contribution from all the previously accreted parcels of gas, $\Mgas(t'),$ for all times $t-\taucons\leq t'\leq t.$ We implement this assumption as follows:
\beq
\label{eq:mssfr}
\\
\sfrtfracms=\mseff(\Mh(t)) \int_{0}^{t}\dtime'\dmqdtpfrac{\Mgas(t')}\cdot\fcons(t\vert t',\taucons)\nonumber.
\eeq
In Eq.~\ref{eq:mssfr}, the gradual transformation of accreted gas into stars is controlled by $\fcons(t\vert t',\taucons),$ which we refer to as the {\em consumption function.} The ``ms" superscript on $\dmqdt{M_{\star}^{\rm ms}}$ in the left-hand side of Eq.~\ref{eq:mssfr} denotes ``main sequence", as this equation refers to the star formation rate that the galaxy would have in the absence of a quenching event (see \S\ref{subsec:wot_quenching} below for our treatment of quenching). Our model for main sequence star formation is therefore parameterized by two separate functions, the baryon conversion efficiency function, $\mseff(\Mhalo(t)),$ defined in the previous section, and $\fcons(t).$ We now define our parameterization for the consumption function.

In modeling the gradual transformation of accreted gas into stars, we make use of the triweight function, $\mathcal{T}(x),$ defined as follows:
\beq
\label{eq:triweight}
\mathcal{T}(x\vert\mu,\sigma) &\equiv&
\begin{cases} 
      0 & y<-3 \\
      z & -3\leq y\leq 3 \\
      0 & y>3
\end{cases}\\
y &=& (x-\mu)/\sigma \nonumber\\
z &=& \frac{35}{96}(1 - (y / 3)^2)^ 3 / \sigma \nonumber.
\eeq
The function $\mathcal{T}(x\vert\mu,\sigma)$ has very similar behavior as a Gaussian centered at $\mu$ with width $\sigma.$ Despite its piecewise definition, the coefficients in Eq.~\ref{eq:triweight} are defined so that $\mathcal{T}(x)$ has continuous derivatives across the real line. Moreover, the function $\mathcal{T}(x)$ can also be evaluated without calls to special functions, making it computationally advantageous for implementations targeting GPUs and other accelerator devices.

The physics captured by the gas consumption function is that freshly accreted gas does not instantaneously turn all its mass into stars, but rather, this transformation is spread out over some timescale, $\taucons,$ that follows the accretion of each new gas parcel. We implement this physical effect in terms of $\mathcal{T}(x)$ as follows:
\beq
\label{eq:fdep}
\fcons(t\ \vert\ \tacc,\taucons) &\equiv&
\begin{cases} 
0 & t<\tacc \\
A\times\mathcal{T}(t\ \vert\ \mu_{\rm cons}, \sigma_{\rm cons}) & t\geq\tacc
\end{cases}\nonumber\\
\mu_{\rm cons} &=& \tacc+\alpha \\
\sigma_{\rm cons} &=& (\taucons-\alpha)/3 \nonumber\\
\alpha &=& \frac{\taucons}{2} \frac{\taucons}{20{\rm \,Gyr}} \nonumber \\ 
A &=& \left(\int_{\tacc}^{\tacc+\taucons} \mathcal{T}(t\ \vert\ \mu_{\rm cons}, \sigma_{\rm cons})\, dt \right)^{-1} \nonumber
\eeq
For a parcel of gas accreted at time $\tacc,$ note that $\taucons$ is the only free parameter that modulates the behavior of $\fcons(t).$ This one-parameter family of functions has a peak at $t=\tacc$ when $\taucons\rightarrow0$, and this peak gradually shifts to later times as $\taucons$ increases.

In Eq.~\ref{eq:fdep}, we have chosen to formulate $\fcons(t)$ in terms of a triweight function, $\mathcal{T}(x),$ that has been normalized to unity when integrated over the time interval $\tacc<t<\tacc+\taucons.$ This formulation has the advantage of mathematically decoupling the roles played by $\mseff$ and $\taucons$ in our model. While this choice comes at the expense of an expression for $\fcons(t)$ that visually appears somewhat complicated, we note that the increase in computational time associated with this choice is practically negligible. 

We show the simple behavior of $\fcons(t)$ in Figure~\ref{fig:depletion}, which illustrates how a parcel of gas accreted at $\tacc$ gradually transforms its mass into stars at later times. Our use of the triweight function ensures that this transformation is fully differentiable, and proceeds to completion over the finite timescale, $\taucons.$ We refer the reader to Appendix~\ref{appendix:sams} for a discussion of the relationship between the \dstar parameter $\taucons$ and the physical interpretation of the consumption timescale in traditional semi-analytic models of galaxy formation.

Figure~\ref{fig:wot_depletion} shows an example fit to a typical early-forming halo with a star formation history that peaks at late times.  The assembly history of this galaxy is significantly better approximated when including the physics of long gas consumption timescales.

\subsection{Quenching and Rejuvenation}
\label{subsec:wot_quenching}

\begin{figure}
\includegraphics[width=8cm]{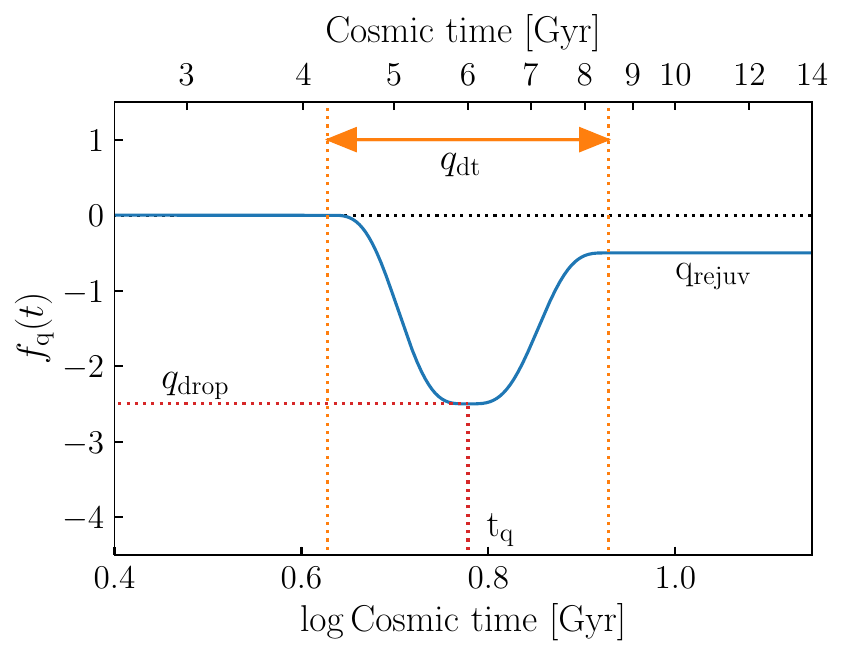}
\caption{{\bf Illustration of the quenching function}. The vertical axis in the figure shows $\qfunc(t),$ the logarithm of the quenching function (see Eqs.~\ref{eq:quenching}-\ref{eq:qfunc}). This function supplies a multiplicative prefactor that shuts down SFR below the main sequence rate. Quenching events can be either rapid or gradual, and our implementation allows for the possibility of a rejuvenation event that returns the galaxy to the main sequence.}
\label{fig:qfunc}
\end{figure}

Observed galaxies present a bimodality in their specific star formation rates (sSFR) and broadband colors \citep{Strateva01, Blanton2003, Baldry2004, Bell2004, wetzel_etal12_persistent_bimodality, Muzzin2013}. Moreover, when galaxy samples are divided according to their color (sSFR), it has been widely found that red (quenched) subsamples reside in higher-density environments relative to blue (star-forming) subsamples \citep{norberg_etal02,blanton_etal05,zehavi_etal05,weinmann_etal06,li_etal06,hearin_etal14_age_matching2}, a phenomenon that persists across across most of cosmic time \citep{coil_etal08,peng_etal10,cooper_etal12}, varies monotonically with color (sSFR) \citep{zehavi_etal11,krause_etal13,coil_etal17,berti_etal21}, and applies to both central and satellite galaxies \citep{wang_etal08,wang_etal13,berti_etal19}. Since galaxies do not traverse cosmological distances greater than $\sim10$ Mpc in a Hubble time, and since the optical colors of a galaxy remain blue for at least $\sim2$ Gyr after the cessation of its star formation \citep[e.g.,][]{conroy_gunn_white_2009}, these observations imply that once star formation in a galaxy has been shut down, the typical galaxy remains quenched.

Of course, not all galaxies are typical, and for a non-negligible minority of galaxies, quenching is not permanent. Numerous observations show that a significant fraction of massive elliptical galaxies have had some recent star formation following a long period of quiescence, a phenomenon generally referred to as {\it rejuvenation} \citep[e.g.][]{Kaviraj2007, Pipino2009, Canning2014, Ehlert2015, Cerulo2019}. Rejuvenation is generally thought to contribute only a small fraction ($<10\%$) of the total stellar mass of a galaxy \citep{Chauke2019}, although observational estimates of the rejuvenation fraction vary considerably (as do the adopted definitions of rejuvenation), ranging from $5\sim30\%$ \citep{Pandya2017,Tacchella2021}. Using forward-modeling techniques based on \um, it was estimated in \citet{behroozi_etal19} that $10-20\%$ of galaxies with stellar mass $\Mstar\gtrsim10^{11}M_\odot$ at $z=1$ (or $40-70\%$ at $z=0$) have experienced some appreciable level of rejuvenation.

In \dstar, we capture these phenomena with our implementation of the quenching function, $\Qfunc(t),$ which acts as a multiplicative factor on the star formation rate:
\beq
\label{eq:quenching}
\sfrtfrac = \Qfunc(t)\times\sfrtfracms.
\eeq
In Eq.~\ref{eq:quenching}, the quantity $\sfrms$ is defined by Eq.~\ref{eq:mssfr}, and we define the behavior of $\Qfunc(t)$ in terms of the logarithmic drop in SFR, $\qfunc\equiv\log\Qfunc,$ which we implement through two successive applications of the triweight error function (cumulative distribution function of the triweight function defined in Eq.~\ref{eq:triweight}), $\terf(x)$:
\beq
\label{eq:triweight_err}
\terf(x\vert\mu,\sigma) &\equiv&
\begin{cases} 
      0 & y<-3 \\
      z & -3\leq y\leq 3 \\
      1 & y>3
\end{cases}\\
y &=& (x-\mu)/\sigma \nonumber\\
z &=&\frac{1}{2} + \frac{35}{96}y - \frac{35}{864}y^3 + \frac{7}{2592}y^5 -\frac{5}{69984}y^7 \nonumber.
\eeq

\beq
\label{eq:qfunc}
\qfunc(\log t &\vert&  \log \qtime,\, \qdt,\,\qdrop,\, \qrejuv) \equiv \nonumber
\begin{cases} 
z_0 & w<0 \\
z_1 & w>0 
\end{cases}\\
w &=& (\log t - \log \qtime) / (\qdt/12)\\
z_0 &=&  \qdrop\times\terf(w+3|\mu=0, \sigma=1)\nonumber\\
z_1 &=& \qdrop - ( \qdrop- \qrejuv)\cdot\terf(w-3|\mu=0, \sigma=1)\nonumber
\eeq

\begin{figure}
\centering
\includegraphics[width=8cm]{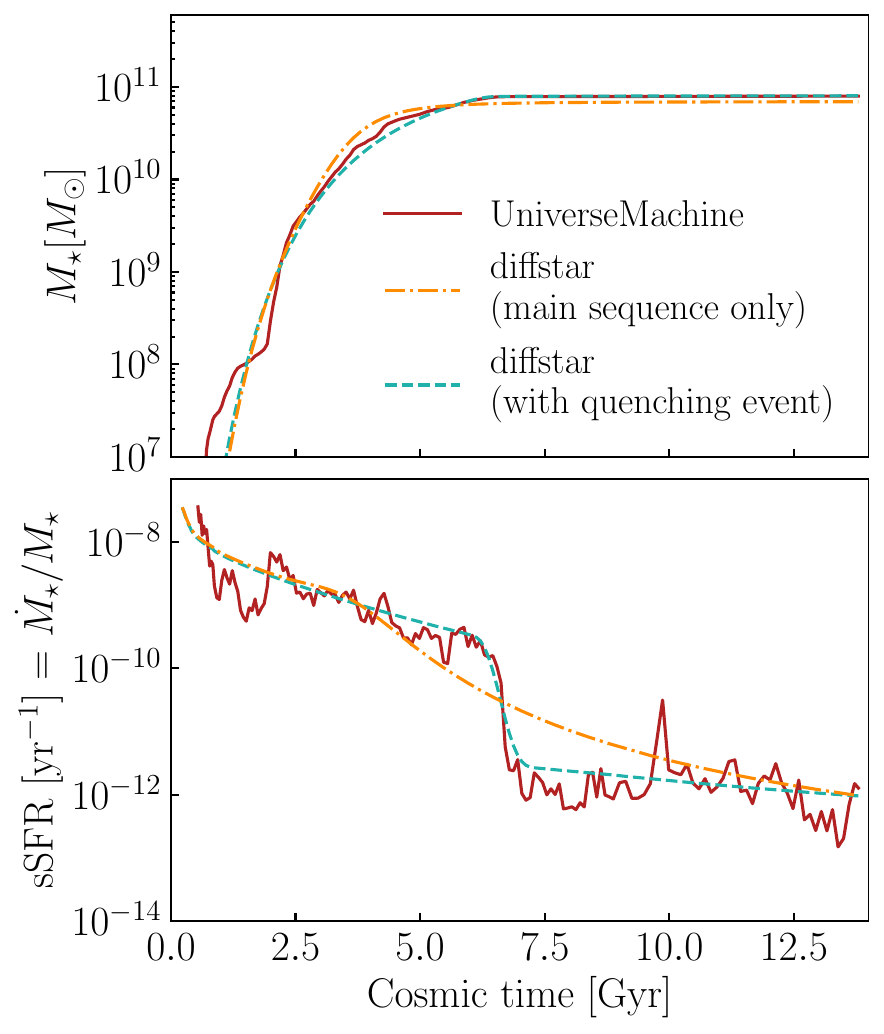}
\caption{{\bf Permanent quenching event}. 
{\it Top panel:}
Stellar mass history. 
{\it Bottom panel:}
Specific star formation rate history ($\mathrm{sSFR}\equiv \dot{M}_\star/M_\star$). 
The solid red lines show the assembly history of a galaxy in \um. The dot-dashed orange lines show an approximation to this history based on the \dstar main sequence model, while the dashed cyan lines show the best-fitting \dstar model that includes the possibility of a quenching event. The implementation of quenching in our model is able to capture a sharp decrease in star formation rate that some galaxies experience when they depart the main sequence.
}
\label{fig:wot_quenching}
\end{figure}

\begin{figure}
\centering
\includegraphics[width=8cm]{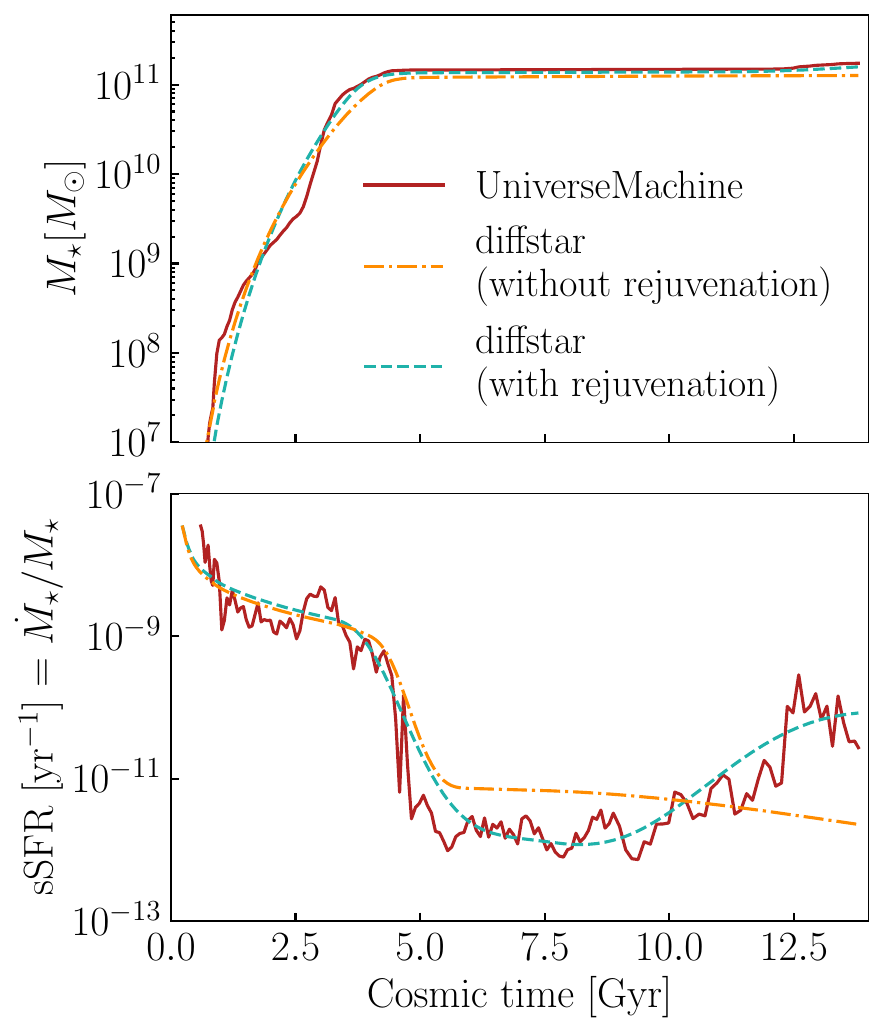}
\caption{{\bf Rejuvenation event}. 
{\it Top panel:}
Stellar mass history. 
{\it Bottom panel:}
Specific star formation rate history ($\mathrm{sSFR}\equiv \dot{M}_\star/M_\star$). 
The solid red lines show the assembly history of a galaxy in \um. The dot-dashed orange lines show an approximation to this history based on a version of \dstar that only permits permanent quenching, while the dashed cyan lines show the best-fitting \dstar model that includes the additional possibility of rejuvenation. The implementation of rejuvenation in our model is able to capture the SFH of a quenched galaxy that resumes a sustained period of star formation, possibly even returning the SFR of the galaxy to the main sequence rate.
}
\label{fig:wot_rejuvenation}
\end{figure}

The parameters in Eq.~\ref{eq:qfunc} are intuitively interpreted as follows: $\qtime$ is the time at which the quenching event reaches $\qdrop,$ its maximum suppression of SFR; the parameter $\qdt$ controls the duration of the quenching event; a quenched galaxy begins to depart the main sequence when the quenching event starts at time $\log \qtime^{\rm start} \equiv \log \qtime - \qdt/2$. The quantity $\qrejuv$ controls the level of rejuvenation; when $\qdrop=\qrejuv,$ the galaxy remains forever quenched; when $\qrejuv=0$, the galaxy eventually returns to the main sequence; finally, note that we require that $\qdrop\leq\qrejuv\leq0,$ so that we do not allow a rejuvenation event to produce SFR in excess of the main sequence rate.
Figure~\ref{fig:qfunc} gives a visual representation of $\qfunc(t),$ and illustrates the physical interpretation of each of the free parameters that regulate its behavior.

In Figure~\ref{fig:wot_quenching}, we show a \dstar fit to a galaxy in \um whose SFH includes a quenching event that is prominent and permanent. The bottom panel shows the \textit{specific} star formation rate history $\mathrm{sSFR}(t)\equiv {\rm SFR}(t) / M_\star(t)$ of a \um galaxy. When we fit the \dstar model to this SFH, our fitter is able to correctly identify the abrupt quenching that reduces SFR by more than two orders of magnitude, finding $t_q\sim 7.2\,{\rm Gyr}$.

Figure~\ref{fig:wot_rejuvenation} shows an example fit to a galaxy in \um that quenched around $t\sim 5\,{\rm Gyr}$, remained quiescent for $\sim 5\,{\rm Gyr}$ and subsequently rejuvenated, ultimately forming $15\%$ of its present-day stellar mass within the last $1.5\,{\rm Gyr}$ of its lifetime. By comparing the dot-dashed orange lines to the dashed cyan lines in Fig.~\ref{fig:wot_rejuvenation}, we can see that the rejuvenation feature built into $\Qfunc(t)$ is a necessary degree of freedom in order for \dstar to capture the SFH of simulated galaxies such as this one.

We conclude this section by calling attention to the relationship in \dstar between the quenching of a galaxy and the growth of its parent dark matter halo. In the absence of a consumption timescale, Eqs.~\ref{eq:dmgasdt} \& \ref{eq:bce} guarantee that once the mass of a halo stops growing, the star formation of its galaxy immediately shuts down, which is inconsistent with the long delay between satellite infall and quenching \citep[e.g.,][]{wetzel_etal13_quenching_timescales,wheeler_etal14_quenching_inefficiency,haines_etal15}. A simple technique to address this shortcoming is to introduce a parameterized delay between $\qtime$ and the time halo growth shuts down; this approach has been used with notable success in the EMERGE model \citep{moster_emerge1,moster_etal20_emerge_passive_galaxies,oleary_etal21_emerge_merger_rates}, although this formulation makes a very specific assumption about quenching that may be difficult to reconcile with the {\it diversity} of quenching pathways \citep[e.g.,][]{fillingham_etal16,balogh_etal16,wright_etal19}. When fitting individual SFHs with the \dstar model, the gas consumption timescale, $\taucons,$ and the quenching timescale, $\qtime,$ are each allowed to vary freely and independently, which ensures that our model is able to capture a considerable diversity in galaxy--halo co-evolution,  permitting both a ``decoupling" between star formation and halo growth in some galaxies, as well as tightly-coupled growth in other galaxy/halo systems.

\section{Diffstar Model Performance}
\label{sec:model_perform}

\begin{figure*}
\centering
\includegraphics[width=\textwidth]{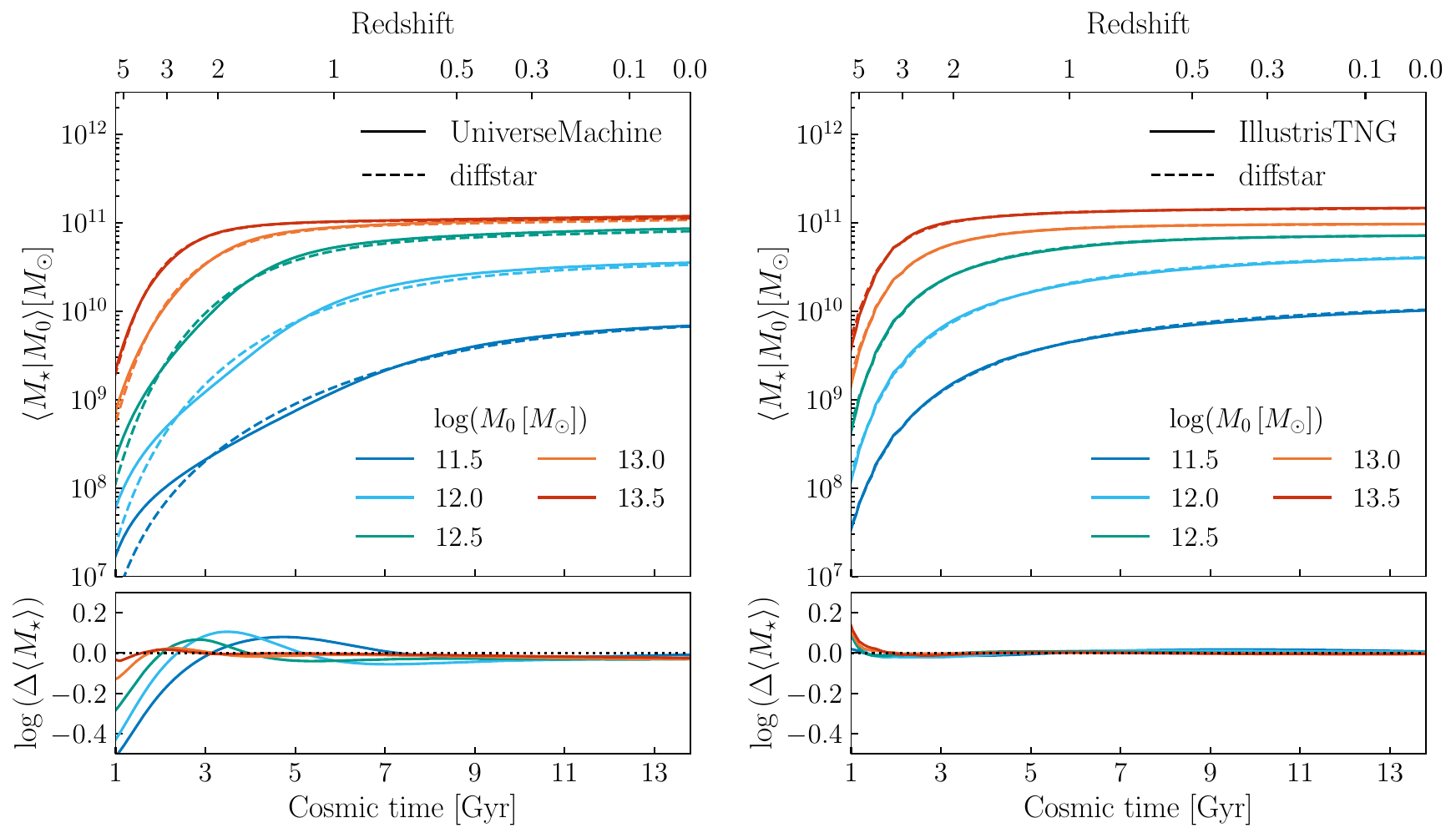}
\caption{{\bf Average stellar mass histories}. 
\textit{Top panels:} Average stellar mass histories (SMH) as a function of cosmic time for halos of different present day mass $M_0\equiv\Mpeak(z=0)$. Dashed curves show the prediction we obtain by fitting \dstar to each individual SMH history and taking the average for the same halos.  \textit{Bottom panels:} Residual logarithmic difference between the average SMH in the simulation and the average prediction from \dstar. We find good agreement between \dstar SMH predictions compared to UM or TNG, especially at $t>3\,{\rm Gyr}\,(z<2)$, with a residual bias of 0.02 dex for TNG and 0.1 dex for UM.
}
\label{fig:average_Mstar}
\end{figure*}

In the previous section, we provided a detailed pedagogical description of the \dstar model for individual galaxy assembly. As we introduced each ingredient of the model, we supplied an illustrative example of a particular simulated SFH whose fit warranted the ingredient under discussion. Since one of our ultimate aims is to deploy our model in a fully cosmological context, a natural question that arises is how well the full diversity of star formation histories in \um and \tng are captured by \dstar. In this section, we quantitatively assess the ability of our model to capture the SFHs seen in these two simulations. In \S\ref{subsec:fitting_procedure}, we describe our algorithm for fitting individual SFHs in simulations with \dstar. We quantify how well our model is able to reproduce {\it average} star formation histories in \S\ref{subsec:average_fits}, we present the residual errors of fits to {\it individual} SFHs in \S\ref{subsec:sfh_residuals}, and we analyze the timescale-dependence of the residual errors of our fits in \S\ref{subsec:fitting_residuals}.

We remind the reader of the notation introduced in \S\ref{sec:sims}, in which ${\rm sSFR}\equiv\log\dot{\rm M}_{\star}/\Mstar\ [{\rm yr^{-1}}],$ and all logarithms are understood to be in base-10.

\subsection{Fitting simulated SFHs with Diffstar}
\label{subsec:fitting_procedure}

The first step to obtaining a \dstar approximation to the SFH in a simulation is to fit the assembly history of the total mass of the halo (i.e., the MAH) with the \dmah model. As discussed in \S\ref{subsec:model_mah}, the \dmah model is specifically formulated to describe the {\it cumulative} MAH, and so these fits are carried out on the simulated history of $\Mpeak(t).$ We adopt the same fitting procedure described in detail in \citet{hearin_etal21_diffmah}, to which we refer the reader for further information. Once a \dmah approximation has been identified, the three best-fitting parameters describing the MAH are held fixed, and we use the smooth approximation to $(M_h(t),\,\dot{M}_h(t))$ in order to approximate the SFH and stellar mass history (SMH) of the galaxy, only varying \dstar parameters in the second stage of the fit.

To fit the \dstar parameters, we use a custom-tailored wrapper\footnote{A fiducial initial guess is selected based on the median value of each parameter as a function of $M_0$ measured from an initial exploratory run. Our fitter reruns the BFGS-based optimization numerous times until a target best-fit loss is obtained, stopping after a maximum number of iterations. Each iteration starts from a different initial guess determined by randomly perturbing the fiducial initial guess.} function calling the {\tt scipy} implementation of the BFGS algorithm \citep{broyden_1970_B_in_BFGS,fletcher_1970_F_in_BFGS,goldfarb_1970_G_in_BFGS,shanno_1970_S_in_BFGS} to carry out a least squares minimization of the logarithmic difference between the model prediction and the target data. For our target data, we use the logarithmic values of SMH, the logarithmic SFH averaged over a time period of $1\,{\rm Gyr}$ (see Equation~\ref{eq:time_smoothed_SFH}), and the specific star formation (SFH/SMH), jointly fitting these three target data vectors for snapshots with $t>1\,{\rm Gyr}$. We generally give each target data vectors the same weight, but we double the weight of: (i) SMH snapshots within 0.1 dex of the present day stellar mass, which highlights quenched snapshots; and of (ii) SFH snapshots within 0.1 dex of the SFH maximum value, which highlights the peak of the SFH data vector. Furthermore, when performing the fits, we clip the simulated and predicted SFHs at a minimum value of ${\rm sSFR}=-12;$ the motivation for this clip is that values of SFR falling below this cutoff are observationally consistent with zero detectable star formation \citep{brinchmann_etal04}, and so we do not penalize a proposed model for a failure in this regime. Finally, we only fit snapshots where the SMH is above $10^7\, [\Msun]$ or where the SMH is within 3.5 dex of the present-day stellar mass.

When varying the free parameters in all our fits, we find that fixing $\beta_{\rm hi}= -1$ does not result in an appreciable loss of accuracy, nor does it increase the magnitude of residual variance of the fits, and so we hold this parameter fixed to these values in all results reported here. We therefore vary a total of 8 parameters for each galaxy: $\theta =\{\beta_{\rm lo},\, \epsilon_{\rm crit},\, \Mcrit,\, \taucons,\,t_q,\, q_{\rm dt},\,q_{\rm drop},\,  q_{\rm rejuv}\}$. The software implementation of our fitting algorithm is included as part of the publicly available \dstarcode source code, to which we refer the reader for additional quotidian details. Our minimization algorithm takes a few hundred CPU-milliseconds per halo.

\subsection{Recovery of average SFH}
\label{subsec:average_fits}

\begin{figure*}
\centering
\includegraphics[width=\textwidth]{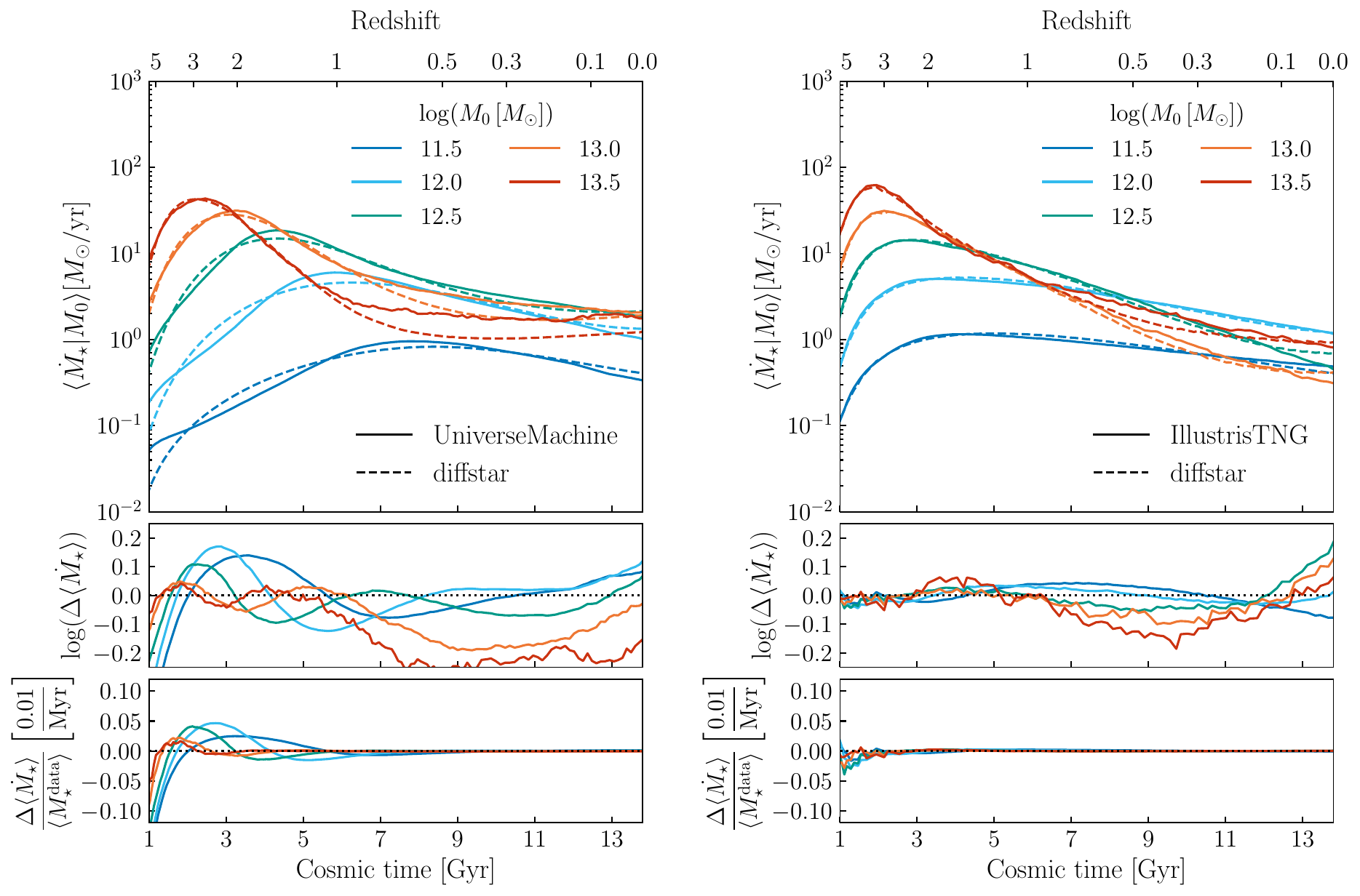}
\caption{{\bf Average star formation histories}. \textit{Top panel}: Star formation histories (SFH). \textit{Middle panel:} Residual logarithmic SFH difference. \textit{Bottom panel:} Residual SFH difference normalized by the simulation SMH in  $[(100\,{\rm Myr})^{-1}]$ units. 
We find good agreement between \dstar SFH predictions compared to UM or TNG, especially at $t>3\,{\rm Gyr}\,(z<2)$, with a residual bias of 0.05 dex for TNG and 0.15 dex UM (except for massive halos at low redshift). Relative to SMH, the SFH residual bias magnitude is typically smaller than $0.01 [(100\,{\rm Myr})^{-1}]$ for TNG and smaller than $0.05 [(100\,{\rm Myr})^{-1}]$ for UM.
}
\label{fig:average_SFH}
\end{figure*}

Using the algorithm described in \S\ref{subsec:fitting_procedure} above, we have identified a best-fitting \dstar approximation to several hundred thousand simulated galaxies in the \um and \tng samples described in \S\ref{sec:sims}. In this section, we analyze how well the {\it average} SMHs and SFHs are described by our model.
Figure~\ref{fig:average_Mstar} shows results for SMHs, showing results for UM in the left column and results for TNG in the right column. Results for galaxies residing in halos with different mass bins are color-coded as indicated in the legend. For galaxies in a particular mass bin, we plot the average SMH with the solid curve in the top panel; solid curves show results taken directly from the simulated merger trees, while dashed curves show results based on \dstar approximations, so that comparing solid to dashed curves illustrates the fidelity with which \dstar approximates the simulated SMH. In the lower panel of Figure~\ref{fig:average_Mstar}, we show the residual logarithmic difference between the average SMH in the simulation and the average prediction from \dstar. We find good agreement between the \dstar SMH predictions compared to UM or TNG, especially at $t>3\,{\rm Gyr}\,(z\lesssim 2)$\footnote{For halos with present-day mass of $m_0=11.5,$ the median mass at $z=3$ is 480 BPL particles, and $10-15\%$ of halos are resolved with fewer than 200 particles; at $z=5,$ the median mass is just over 40 particles.}, with a residual average bias of 0.02 dex for TNG and 0.1 dex for UM. Figure~\ref{fig:average_SFH} has a similar layout as Figure~\ref{fig:average_Mstar}, with the top panel showing average SFHs for simulated halos with solid lines and \dstar approximations with dashed lines. The middle panel shows the residual logarithmic SFH difference, while the additional bottom panel shows the residual SFH difference relative to the simulation SMH in $[(100\,{\rm Myr})^{-1}]$ units. We find good agreement between the \dstar SFH predictions compared to UM or TNG, especially at $t>3\,{\rm Gyr}\,(z\lesssim2)$, with a residual bias of 0.05 dex for TNG and 0.15 dex UM (except for massive halos at low redshift). Relative to the SMH, the SFH residual bias magnitude is typically smaller than $0.01 [(100\,{\rm Myr})^{-1}]$ for TNG and smaller than $0.05 [(100\,{\rm Myr})^{-1}]$ for UM. Figure~\ref{fig:average_Mstar}-\ref{fig:average_SFH} demonstrate that \dstar is a flexible enough model to approximate average galaxy growth in both UM and TNG with a high level of accuracy.

\subsection{Residuals of fits to individual SFH}
\label{subsec:sfh_residuals}

\begin{figure*}
\centering
\includegraphics[width=0.9\textwidth]{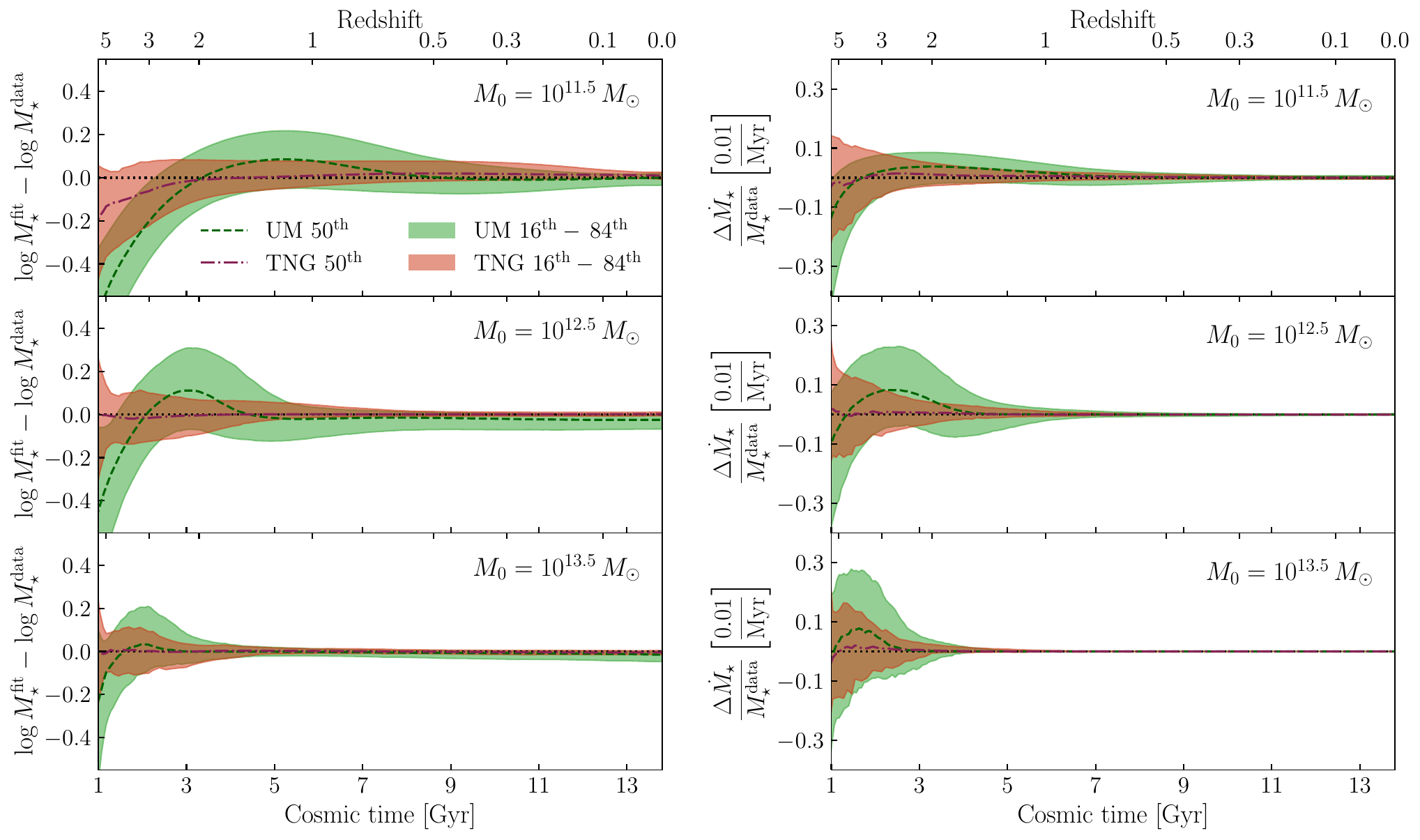}
\caption{{\bf Residuals of fits to individual SFH.} The distribution of residual errors in individual fits from \dstar to UM and TNG as a function of time. {\it Left columns:} residual errors in stellar mass histories (SMH); {\it Right columns:} residual errors in SFHs normalized by the SMHs from the simulation, in units of $[(100\,{\rm Myr})^{-1}]$. Lines show the median of the distribution, while bands show the area between the $16^{\rm th}$ and $84^{\rm th}$ percentiles, representing the residual variance of individual fits. Each panel shows results for halos of different present day mass as indicated by the in-panel annotation. We recover unbiased histories for $t\gtrsim 3\,{\rm Gyr}$ ($z\lesssim 2$), finding a typical residual variance of 0.1 dex or lower for SMH and of $0.05 [(100\,{\rm Myr})^{-1}]$ and lower for SFH differences relative to SMH. The width of these residual distributions decreases significantly at lower redshift, especially as galaxies become quiescent. See \S\ref{subsec:sfh_residuals} for details.}
\label{fig:residuals_sfh}
\end{figure*}

The results shown in \S\ref{subsec:average_fits} illustrate the accuracy with which the \dstar model is able to reproduce the {\it average} assembly history of galaxies in \um and \tng. In this section, we study how faithfully the model can capture the {\it diversity} of SFHs of simulated galaxies, and so we now assess the performance of \dstar in reproducing the assembly history of individual galaxies.

In the left column of Figure~\ref{fig:residuals_sfh}, we show the distribution of residual errors of the \dstar model fits to the SMH of individual galaxies in UM and TNG. The vertical axis shows the logarithmic difference between the simulated and approximated SMH, plotted as a function of cosmic time. The median bias for UM (TNG) is shown with the dashed green (dot-dashed red) curve, and the variance of the logarithmic difference is shown with the shaded band of the corresponding color, defined as the area between the $16^{\rm th}$ and $84^{\rm th}$ percentiles of the distribution of the residual variance. Results for galaxy samples residing in halos of different mass are shown in different panels, with mass range indicated by the in-panel annotation.

Generally speaking, the SMH fits have negligible bias for all times $t\gtrsim3\,{\rm Gyr}$ and for all halo masses studied here, and present a total residual variance of around 0.1 dex or lower (width of the bands). For the case of TNG, SMH fits to galaxies of all mass retain this same level of quality at all redshifts $z\lesssim5.$  For the case of UM galaxies in halos with present-day mass $M_0\approx10^{11.5}\Msun$ (the lowest mass bin we study), at early times there is a systematic offset of 0.2 dex at $z\approx3$ that grows to 0.4 dex at $z\approx5;$ this offset is reduced and pushed to higher redshift for more massive halos in UM, and it generally stays to levels below 0.1 dex for most times in UM halos with $M_0\gtrsim10^{12}\Msun.$ It is plausible that the resolution limits of the underlying BPL simulation could contribute significantly to this offset, but a dedicated resolution study would be required in order to quantify the extent to which this is the case; we discuss this issue further in \S\ref{sec:discussion}. The SMH residual variance is typically lower for TNG than for UM, which is largely attributable to the greater degree of burstiness in UM (see \S\ref{subsec:fitting_residuals} for further details). Furthermore, the width of the residual variance decreases for more massive halos, which can be understood in terms the increased quenched fraction at higher mass.

In the right column of Figure~\ref{fig:residuals_sfh}, we show analogous results for the ability of our model to describe the SFH of individual galaxies. To quantify these residuals, we adopt a convention similar to \citet{lower_etal20} and plot the difference between simulated and best-fitting approximations of SFHs, normalizing this difference by the SMH in the simulation in $[(100\,{\rm Myr})^{-1}]$ units. The vertical axis quantifies the residual error in the {\it specific} star formation rate \citep[see, e.g.,][for discussion of the relationship between this quantity and the ability of a model to recover galaxy colors]{chaves_montero_hearin_2020_sbu1}. Again we find that the \dstar approximations have a negligible bias for all times $t\gtrsim3\,{\rm Gyr}$ and for all halo masses; for the case of UM galaxies there is an offset of $\sim0.1 \,(100{\rm Myr})^{-1}$ at $z\gtrsim 2 $, but otherwise biases in both simulations are limited to levels below $0.05 \,(100{\rm Myr})^{-1}$ at all times and for all halo masses we consider. The residual variance in this quantity is typically lower than $0.05 \,(100{\rm Myr})^{-1}$ for $z<2$, becoming significantly smaller at lower redshift. The results plotted here can be compared to Figure 5 of \citet{lower_etal20}, where they constrain SFH from synthetic broadband photometry. Evidently, the typical error in our fits is smaller than the precision of typical SPS codes at inferring SFH from galaxy photometry.

\subsection{Residuals from short-timescale fluctuations}
\label{subsec:fitting_residuals}

\begin{figure*}
\centering
\includegraphics[width=0.9\textwidth]{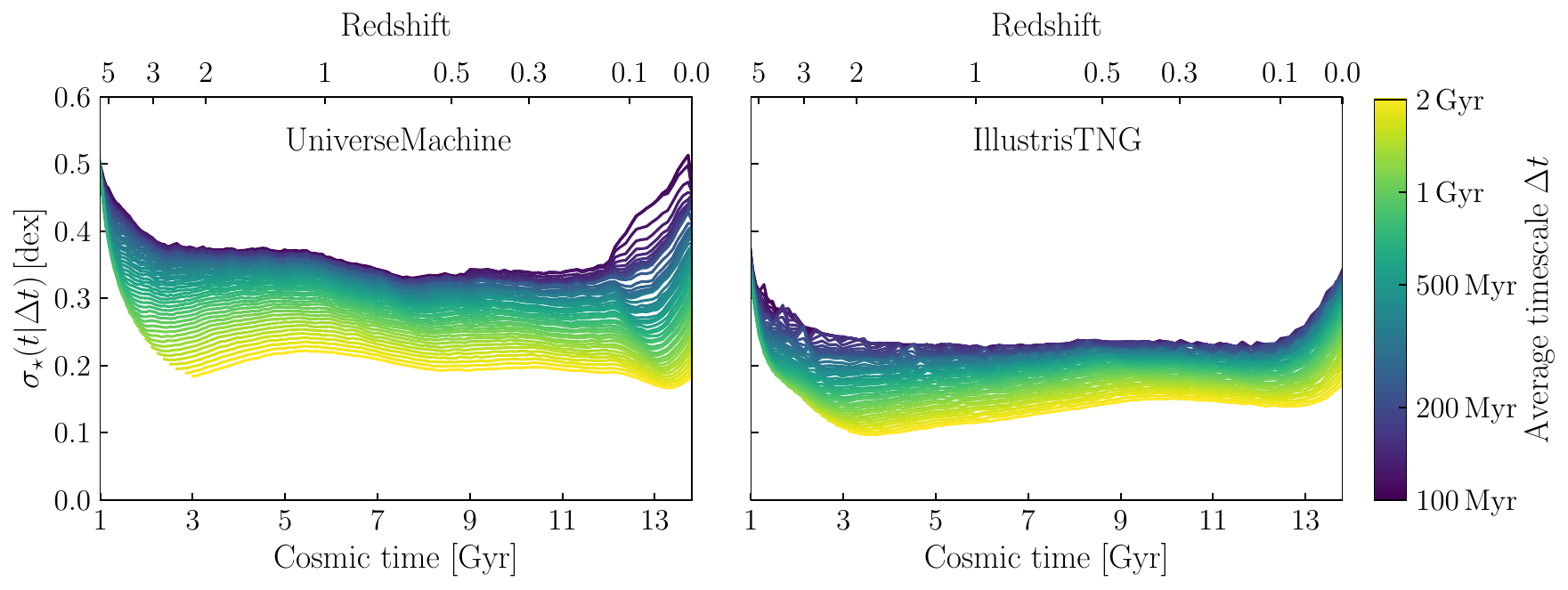}
\caption{{\bf Residual errors from short-timescale fluctuations}. Both panels show the residual variance in the \dstar approximations to SFH, $\sigma_{\star}(t\vert\Delta t),$ as a function of time; variance arising from transient fluctuations on different timescales $\Delta t$ are plotted with different colored curves, as indicated by the color bar. Larger values of the y-axis correspond to a greater degree of burstiness in the simulations that is not captured by the \dstar approximation. See Eq.~\ref{eq:time_smoothed_SFH}-\ref{eq:time_smoothed_SFH_variance} for the definition of $\sigma_\star(t\vert\Delta t).$ Results for UM are shown in the left panel, and TNG in the right panel. Relative to UM, the SFH histories in TNG  are smoother, and are approximated with lower variance by \dstar.
}
\label{fig:residual_SFR_variance}
\end{figure*}

Star formation histories of individual galaxies in simulations fluctuate on shorter time scales than can be described by the smooth \dstar model. In this section, we explore the extent to which these transient fluctuations are responsible for residual variance in the \dstar approximations to the SFHs of individual galaxies in simulations.

We begin by defining how we quantify the difference between simulated SFHs and their \dstar approximations on a particular timescale, $\Delta t.$ First, we use the notation $\langle \dot{M}_\star(t) \rangle_{\Delta t},$ referred to as the $\Delta t$-smoothed SFH, to quantify the amount of star formation that has occurred over the timescale $\Delta t$ prior to the time $t:$
\begin{equation}
\label{eq:time_smoothed_SFH}
\langle \dot{M}_\star(t) \rangle_{\Delta t} \equiv (M_\star(t) - M_\star(t-\Delta t)) / \Delta t.
\end{equation}
Second, for a particular galaxy, we use the notation $\delta\langle \dot{M}_\star(t) \rangle_{\Delta t}$ to denote the logarithmic difference between the $\Delta t$-smoothed SFH in the simulation and its \dstar approximation:
\beq
\label{eq:time_smoothed_SFH_error}
\\
\delta\langle \dot{M}_\star(t) \rangle_{\Delta t}\equiv \log\left( \langle \dot{M}_\star(t) \rangle_{\Delta t}^{\rm fit}\right) -  \log\left( \langle \dot{M}_\star(t) \rangle_{\Delta t}^{\rm sim}\right)\nonumber.
\eeq
Finally, for some probability distribution $P(x),$ we use the notation $\sigma_{68}$ to refer to the half-width between the $16{\rm th}$ and $84{\rm th}$ percentiles of $x.$ 

Based on the three quantities defined above, we can quantify the fidelity with which the \dstar model captures simulated SFHs on a timescale $\Delta t$ for any particular sample of galaxies. For each galaxy in the sample of interest, we compute $\delta\langle \dot{M}_\star(t) \rangle_{\Delta t},$ and then define $\sigma_\star(t\vert\Delta t)$ to be the value of $\sigma_{68}$ for the resulting distribution:
\beq
\label{eq:time_smoothed_SFH_variance}
\sigma_\star(t\vert\Delta t) \equiv \sigma_{68}\left( \delta\langle \dot{M}_\star(t) \rangle_{\Delta t} \right).
\eeq
We use $\sigma_\star(t\vert\Delta t)$ as our metric to assess the time-scale dependence of the success of the \dstar model. 

In Figure~\ref{fig:residual_SFR_variance}, we plot $\sigma_\star(t\vert\Delta t)$ as a function of time, showing results for all \um galaxies in the left panel, and all galaxies in \tng in the right panel. Results for different smoothing scales $\Delta t$ are color-coded as indicated in the color bar. At $\Delta t=0$, corresponding to the smallest timescale resolved by the snapshot spacing of the simulated datasets (see \S\ref{sec:sims}), we find a typical residual variance of 0.35 dex for UM, and 0.25 dex for TNG, comparable to a typical observational error on the value of SFR inferred from galaxy spectra \citep[e.g.,][]{brinchmann_etal04}. We find that \um histories present a greater degree of burstiness relative to \tng, confirming previous results \citep{iyer_etal20,chaves_montero_hearin_2021_sbu2}. As \dstar is a smooth parametric model, we generally expect better predictions for SFH when averaged over longer timescales. This expectation is borne out quantitatively, as Figure~\ref{fig:residual_SFR_variance} shows that $\sigma_\star$ decreases monotonically with increasing values of $\Delta t.$ When the SFH is smoothed over a time period of $2\,\rm{Gyr}$, \dstar captures SFH within 0.2 dex for UM, and 0.15 dex for TNG. In \S\ref{sec:discussion}, we discuss our ongoing work in developing an extension of \dstar that incorporates the short-timescale fluctuations that give rise to these residuals.

\section{Interpreting Simulations with Diffstar}
\label{sec:interpretaish}

The parameters of the \dstar model have simple interpretations in terms of key scaling relations that emerge from the physics of galaxy formation. In this section, we study the statistical distribution of the parameters of the best-fitting approximations to the SFHs in \um and \tng, and show how comparing these distributions gives insight into the similarities and differences between these two simulations in terms of the basic physical picture offered by our model.

We remind the reader of the notation introduced in \S\ref{sec:sims} in which we use the variable $m_{\rm h}\equiv\log\Mh [\Msun]$, and ${\rm sSFR}\equiv\log\dot{\rm M}_{\star}/\Mstar [{\rm yr}^{-1}],$ with all logarithms understood to be in base-10.

\subsection{Main Sequence Star Formation}
\label{subsec:ms_comparison}

Main sequence galaxies in the \dstar model only ever convert a fraction $\mseff$ of their accreted gas into stellar mass; as discussed in \S\ref{subsec:wot_bce}, we refer to this fraction as the {\it baryon conversion efficiency.} Additionally, in \dstar we assume that the conversion of accreted gas into stars is a gradual process that takes place over the {\it gas consumption timescale, $\taucons.$} These two ingredients form the basis of the \dstar model of main sequence star formation. In this section, we use the best-fitting \dstar approximations to the simulated SFHs presented in \S\ref{sec:model_perform} to compare the UM and TNG models for main sequence galaxies, showing results pertaining to $\mseff$ in \S\ref{subsubsec:bce_comparison}, and to $\taucons$ in \S\ref{subsubsec:taucons_comparison}. 

\subsubsection{Baryon conversion efficiency}
\label{subsubsec:bce_comparison}

\begin{figure}
\centering
\includegraphics[width=8cm]{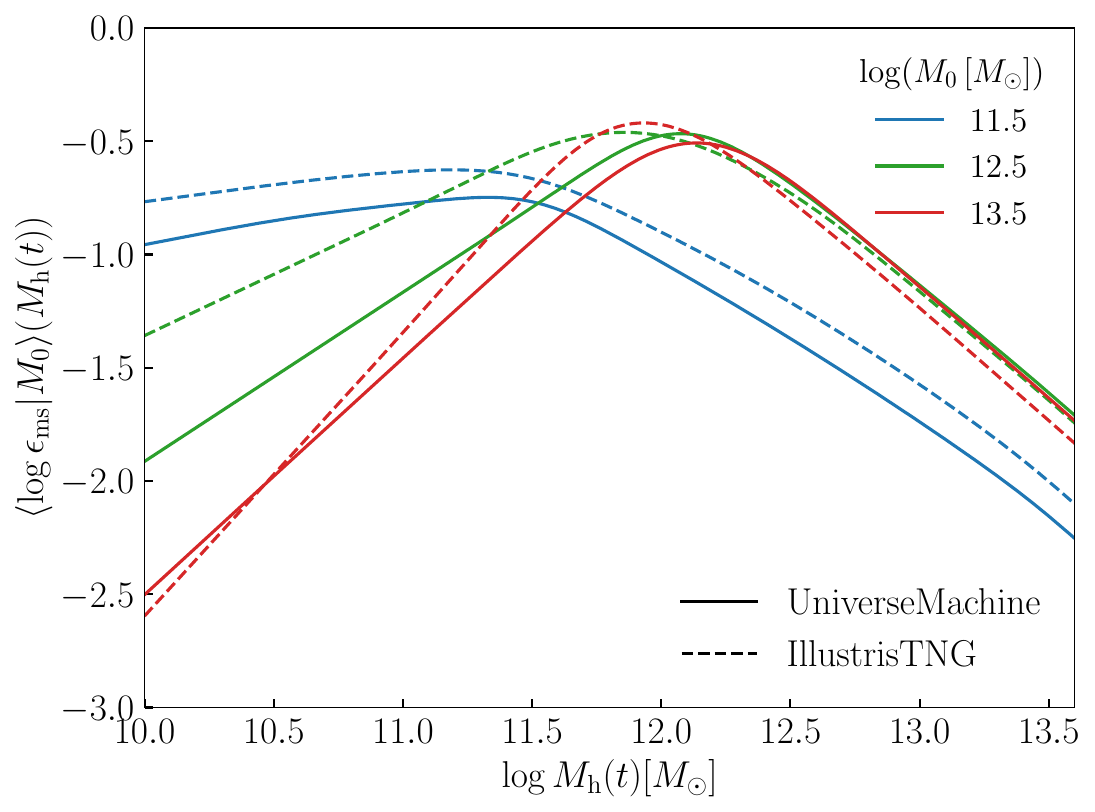}
\caption{{\bf Main Sequence efficiency.}
Average baryon conversion efficiency of main sequence galaxies in TNG and UM (see Equation~\ref{eq:mseff}). Each curve is computed from the collection of best-fitting \dstar approximations to simulated galaxies residing in halos in a narrow bin of present-day mass. Main sequence galaxies in TNG generally convert a larger fraction of their accreted gas into stars relative to UM.
}
\label{fig:efficiency_results}
\end{figure}

As described in \S\ref{subsec:wot_bce}, an ansatz of our model is that the baryon conversion efficiency is determined only by instantaneous halo mass, $\mseff(\Mhalo),$ and has a characteristic shape summarized in Figure~\ref{fig:sfreff}. When fitting the SFH of each individual galaxy in a simulation, we allow the parameters of $\mseff$ to vary freely as part of the \dstar approximation to its assembly history (see \S\ref{subsec:fitting_procedure} for details about our fitting procedure). 

In Figure~\ref{fig:efficiency_results}, we plot $\mseff$ as a function of halo mass for galaxies in the \um and \tng simulations. We show the conversion efficiency of galaxies residing in halos of different present-day mass with different colored curves as indicated in the legend; solid curves show results for galaxies in UM, and dashed curves show results for TNG. To calculate each curve in the figure, we used the best-fitting \dstar approximations to the simulated SFHs, and at each value of halo mass plotted on the x-axis of Fig.~\ref{fig:efficiency_results}, we computed the average value of $\log \mseff(\Mhalo)$ for the galaxies in the mass bin.

Broadly speaking, Figure~\ref{fig:efficiency_results} shows that $\mseff$ tends to be larger in TNG relative to UM, particularly at low mass; this tells us that an accreted parcel of gas in TNG tends to form more stellar mass than in UM, i.e., star formation in TNG is more efficient than in UM.\footnote{Note that the consumption function has no effect on the total mass formed from an accreted parcel of gas, as $\int_{t'}^{t'+\taucons} \fcons(t\vert t',\taucons)\,dt=1$. In principle, this choice of normalization decouples the influence of $\taucons$ from $\mseff,$ since the latter determines the total stellar mass formed from an incoming parcel of gas, and the former controls the timescale over which the transformation takes place. However, in practice, if the consumption time is sufficiently long, then the transformation of gas into stars may not have terminated by $z=0,$ which in effect leads to a degeneracy between $\taucons$ and $\mseff$ when only considering the SFH up until the present day. As we will see in the next section, our conclusions regarding the relative star formation efficiency of TNG vs. UM are not impacted by this degeneracy.}$^{,}$\footnote{While the dashed curves are generally above the solid in Fig.~\ref{fig:efficiency_results}, we can see from the red curves that the efficiency in UM is slightly greater than in TNG at large values of $\Mhalo(t)$ for galaxies residing in halos with $m_0=13.5.$ However, by the time such halos reach large values of $\Mhalo(t)$ in their history, their galaxies tend to be quenched, and so the slight differences in $\mseff$ are immaterial.} In both simulations, $\mseff(\Mhalo)$ peaks at larger values of $\Mhalo$ for galaxies living in halos with larger $m_0.$ In particular, for galaxies residing in massive present-day halos ($m_0\sim13.5$), $\mseff$ peaks at $m_{\rm crit} \sim 12.1$ in UM, and at $m_{\rm crit} \sim 12.0$ in TNG; as $m_0$ decreases, the peak in $\mseff$ gradually and monotonically shifts to lower values, reaching $m_{\rm crit} \sim 11.4$ in UM, and at $m_{\rm crit} \sim 11.2$ in TNG for galaxies in lower-mass halos ($m_0\sim11.5$). These trends indicate that the efficiency with which accreted gas is converted to stars is strongly correlated with the peak in the initial density field from which the galaxy forms, and that at fixed instantaneous halo mass, there is a significant diversity in baryon conversion efficiency.

\subsubsection{Gas consumption timescales}
\label{subsubsec:taucons_comparison}

\begin{figure}
\centering
\includegraphics[width=8cm]{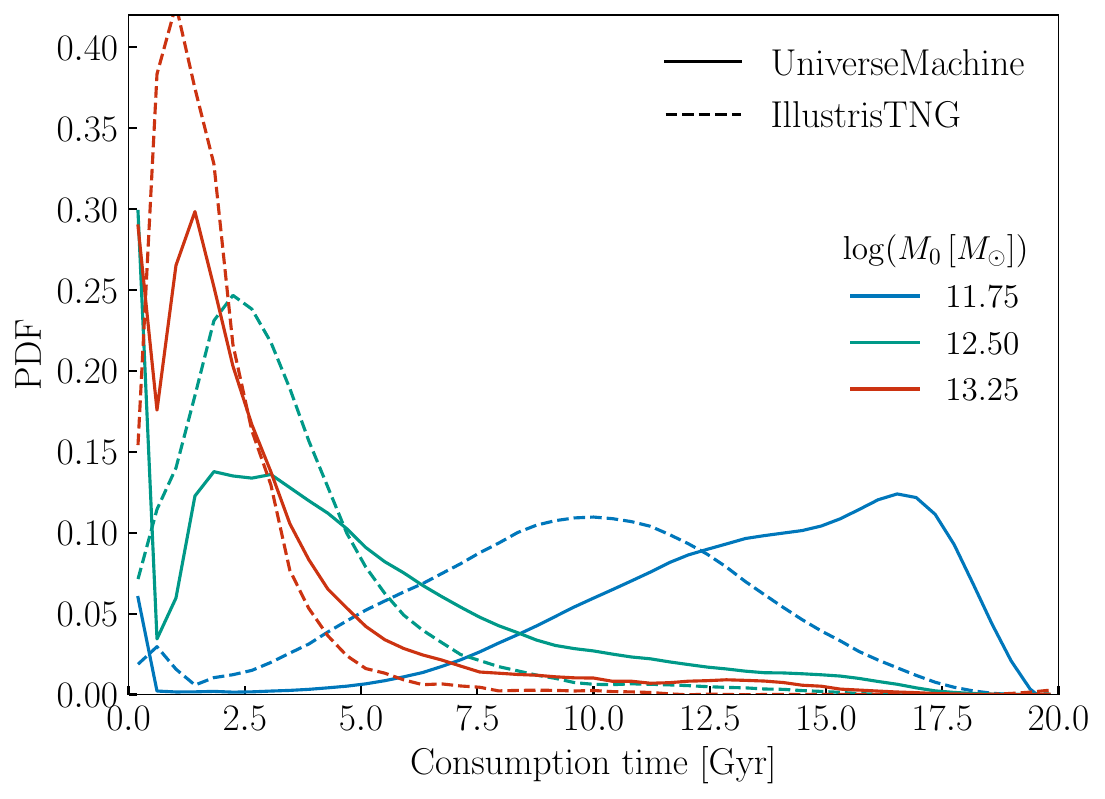}
\caption{{\bf Gas consumption timescales.} 
The distribution of consumption timescales, $\taucons,$ for galaxies in TNG and UM. Galaxies residing in more massive halos have a narrow distribution peaked at $\taucons<3\,{\rm Gyr}$. Galaxies in lower mass halos have a broad distribution peaked at $\taucons\sim 10\,{\rm Gyr}$ in TNG, and peaked at $\taucons\sim 17\,{\rm Gyr}$ in UM.
}
\label{fig:depletion_times}
\end{figure}

As described in \S\ref{subsec:wot_depletion}, when fitting individual SFHs in simulations with the \dstar model, we allow the value of the gas consumption timescale, $\taucons,$ to vary as a free parameter in the fit. In Figure~\ref{fig:depletion_times}, we show the probability distributions of $\taucons$ for simulated galaxies in UM (solid curves), and in TNG (dashed curves). We show the consumption timescales of galaxies residing in halos of different $m_0$ with curves that are color-coded as indicated in the legend.

The gas consumption timescales of galaxies in UM tends to be larger relative to galaxies in TNG. In both simulations, there is a strong dependence of $P(\taucons)$ upon present-day halo mass, $m_0$: galaxies residing in halos with large $m_0$ present a narrow distribution that peaks at $\taucons\sim 0-3\ {\rm Gyr}$, while  galaxies in lower-mass halos have a much greater diversity of consumption timescales that peaks at $\taucons\sim 10\ {\rm Gyr}$ in TNG and at $\taucons\sim 17\ {\rm Gyr}$ in UM. This further supports that TNG is more efficient, as the transformation from gas to stars happens in a shorter, more concentrated timescale. 

\begin{figure}
\centering
\includegraphics[width=8cm]{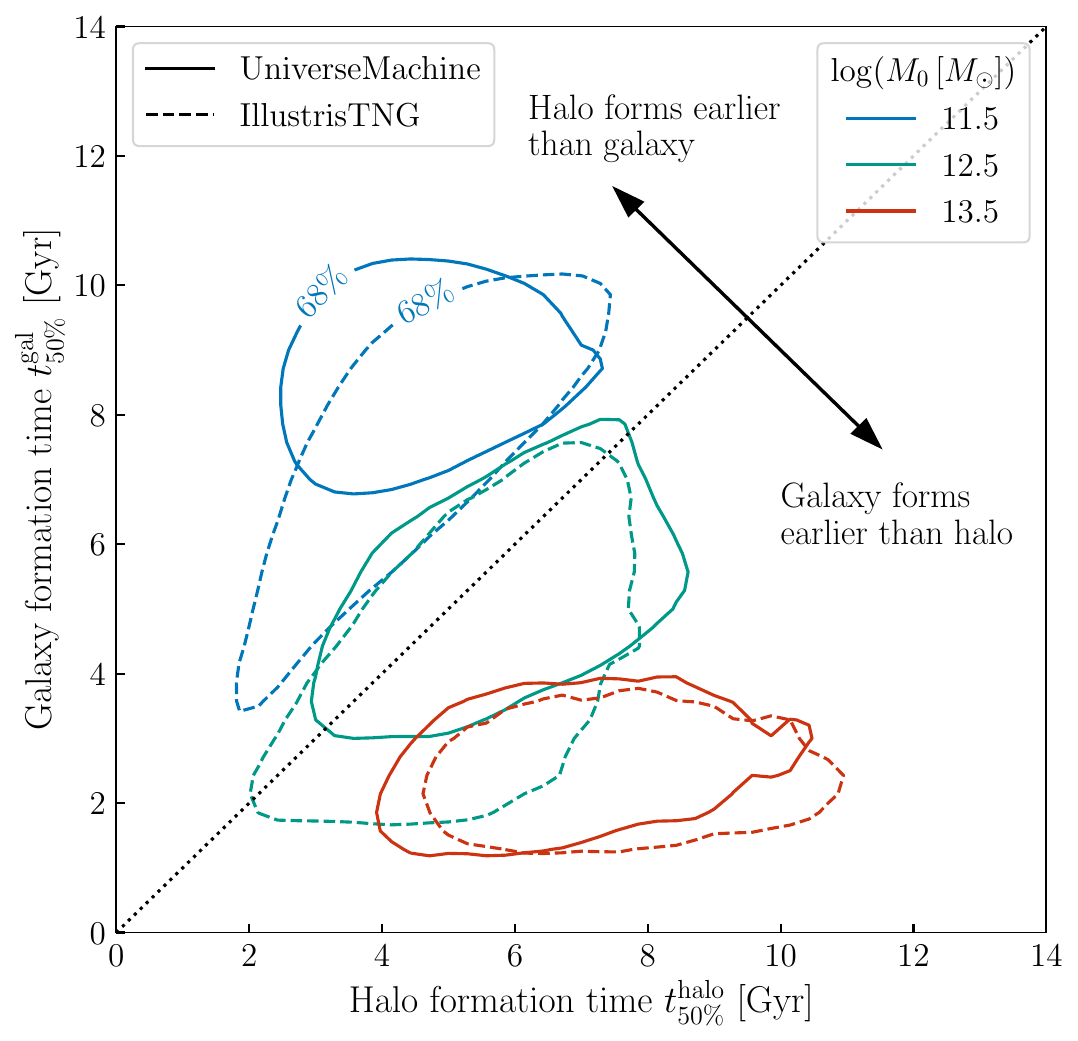}
\caption{{\bf Formation times of galaxies and halos.} Statistical distribution of formation time, defined as the first time the galaxy/halo attained half of its present-day mass, $t_{\rm 50\%}$. Each contour encloses the region where $68\%$ of the objects in the mass bin are found. The dotted line marks the one-to-one relation: for objects lying in the top-left, the halo forms earlier than the galaxy it hosts, and vice-versa for objects in the bottom-right. Lower mass halos with $m_0=11.5$ tend to form earlier than the galaxy they host; this trend reverses for galaxies in more massive halos with $m_0=13.5.$ These trends are closely connected to the mass-dependence of star formation efficiency and gas consumption timescales shown in Figures~\ref{fig:efficiency_results}-\ref{fig:depletion_times}.
}
\label{fig:formation_times}
\end{figure}

We can gain further insight into this trend from Figure~\ref{fig:formation_times}, which compares the formation time of each galaxy to the formation time of its parent halo. To quantify the formation time of an object, we use $t_{50\%},$ the time at which the mass of the galaxy/halo first exceeds half of its peak mass at $z=0.$ In Figure~\ref{fig:formation_times}, on the vertical axis we show the formation time of the galaxy, $t^{\rm gal}_{50\%},$ and on the horizontal axis we show the formation time of the halo, $t^{\rm halo}_{50\%};$ thus as indicated by the annotated arrow, the upper-left quadrant corresponds to halos that form earlier than the galaxies they host, and conversely for the lower-right quadrant. Each ellipsoid in the figure encloses the $68\%$ contour of the typical galaxies in the simulation; contours for UM are shown with solid curves, and contours for TNG are plotted with dashed curves; results for galaxies residing in halos of different $m_0$ are color-coded as indicated in the legend.

We can see clearly from Figure~\ref{fig:formation_times} that in both TNG and UM, low-mass halos tend to form earlier than the galaxies they host, and that this trend is reversed in massive halos, in which the galaxies form earlier than their parent halo.  This is a form of the well-known phenomenon of {\it cosmic downsizing} \citep[e.g.,][]{cowie_etal96,brinchmann_ellis2000,juneau_etal05}. As discussed in \citet{conroy_wechsler_2009}, this terminology has been used in the literature to refer to a wide range of observational phenomena, and so in the present context, we refer to the form of downsizing on display in Figure~\ref{fig:formation_times} as {\it galaxy/halo growth inversion}, by which we mean that low-mass galaxies tend to form later than their parent halos, while high-mass galaxies form earlier than their halos. 

Inverted galaxy/halo growth can be understood in terms of the $\Mhalo$-dependence of the two basic physical ingredients in the \dstar model for main sequence galaxies: $\mseff$ and $\taucons.$ First, due to the general shape of $\mseff(\Mhalo),$ the bulk of star formation in the universe occurs in galaxies residing in dark matter halos with masses $\Mhalo(t)\approx M_{\rm crit}\approx10^{12}M_{\odot}$ \citep[see, e.g.,][]{behroozi_etal2013b}. This implies that massive galaxies residing in dark matter halos with $M_0>M_{\rm crit}$ will tend to form more of their stellar mass at earlier times when their star formation is more efficient; on the other hand, the star formation efficiency of galaxies in lower-mass halos with $M_0\lesssim M_{\rm crit}$ increases monotonically for most or all of cosmic time, and so these galaxies will tend to form a larger proportion of their stellar mass at later times. 

Second, cosmological populations of low-mass dark matter halos present a broad diversity of assembly times, and many of these halos form via assembly histories with $\dmqdt{\Mhalo(t)}$ that declines rapidly following $t^{\rm halo}_{50\%}\approx3$ Gyr, even though the resident galaxies of these halos are still actively assembling in-situ stellar mass today; this indicates that $\taucons\gtrsim10$ Gyr for galaxies in these halos, since the bulk of the gas fueling their ongoing star formation was accreted long ago. Meanwhile, most massive galaxies experience the majority of their in-situ mass growth at redshifts $z\gtrsim2,$ which in the \dstar model is only possible if $\taucons\lesssim2$ Gyr. Thus Figures~\ref{fig:efficiency_results}-\ref{fig:formation_times} taken together make it clear that the inversion of galaxy/halo growth is a basic consequence of the $\Mhalo$-dependence of both $\mseff$ and $\taucons.$ We conjecture that {\it any} observationally successful model of galaxy formation should result in gas consumption timescales with an $\Mhalo$-dependence that closely resembles the behavior of the distributions shown in Fig.~\ref{fig:depletion_times}, and we predict that the general behavior of $\mseff(\Mhalo)$ and $\taucons(\Mhalo)$ shown here will be directly confirmed by future analyses in which of the \dstar model is used to fit the observed SEDs of large samples of galaxies (see \S\ref{sec:discussion} for further discussion).

\subsection{Quenching Time Distributions}
\label{subsec:quenching_comparison}

\begin{figure}
\centering
\includegraphics[width=8cm]{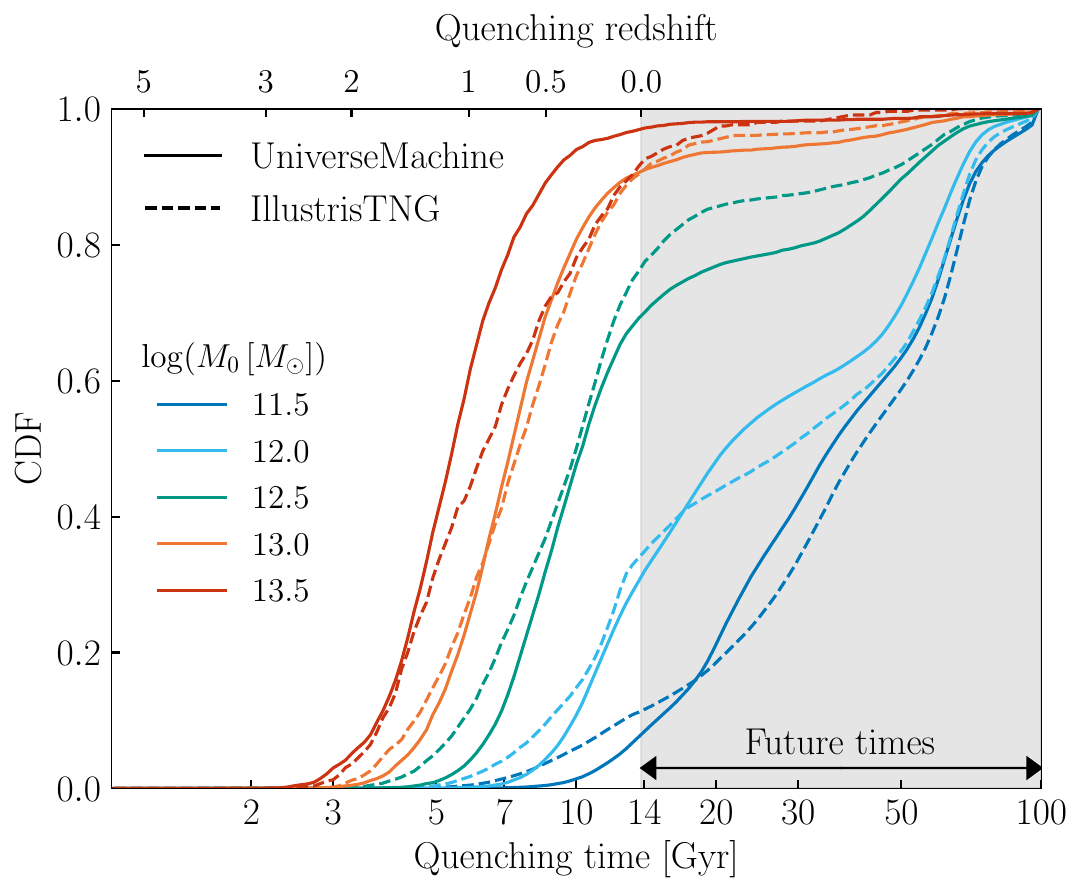}
\caption{{\bf Quenching times.} Each curve shows the fraction of objects that quenched earlier than time $\qtime$ on the x-axis. Results for different mass bins are color-coded according to the legend, with TNG shown with dashed curves, and UM with solid curves. In both simulations, galaxies in more massive halos are more likely to quench at earlier times. Most galaxies in  halos with $\log M_0 > 13.5$ have quenched by present-day, $t_0$, while most galaxies in halos with $\log M_0 < 11.5$ are still on the main sequence. For the case of massive halos, galaxies in TNG have a broader diversity of quenching times that peaks at lower redshift relative to UM. The gray shaded region shows quenching times with $\qtime \gtrsim t_0,$ corresponding to quenching events that have little or no effect on the SFH of observable galaxies, although the quantitative impact also depends on the value of the quenching speed, $\qdt$ (see Figure~\ref{fig:qfunc}).
}
\label{fig:quenching_times}
\end{figure}

When fitting the SFHs of individual galaxies in simulations, the \dstar model for quenching has four free parameters: the quenching time, $\qtime,$ the quenching speed, $\qdt,$ the magnitude of the quenching event, $\qdrop,$ and the magnitude of rejuvenation, $\qrejuv$ (see Fig.~\ref{fig:qfunc}). Quenching in \dstar is not a binary phenomenon that is either ``on" or ``off", but rather, the SFH of a galaxy is jointly fit with all eight of the free parameters of the model, and the best-fitting values of the four quenching parameters capture the extent to which a sustained quenching event plays a significant role in the assembly history of the galaxy.

In Figure~\ref{fig:quenching_times}, we plot the cumulative probability distribution of the quenching time, $\qtime,$ for galaxies in \um (solid curves) and \tng (dashed curves); results for galaxies in halos of different $m_0$ are color coded as indicated in the legend. The shaded region represents future times $t>t_0$, and quenching times with $\qtime \gg t_0$ have little or no effect on the SFH of observable galaxies. Galaxies in both simulations show the same qualitative trend of $\qtime$ with $m_0:$ most galaxies in massive halos ($m_0>13.5$) have experienced a quenching event at some point prior to the present day, $t_0,$ while most galaxies in smaller halos ($m_0<11.5$) have $\qtime\gtrsim t_0,$ and thus remain on the main sequence. In both simulations, as $m_0$ increases the fraction of galaxies that have experienced a quenching event gradually increases, and the distribution of quenching times, $P(\qtime),$ gradually shifts towards lower values of $\qtime$, so that galaxies in more massive halos tend to quench at earlier times.

Even though quenching in \dstar is not a binary designation, we can nonetheless compare $\qtime$ to $t_0$ as a useful criterion to assess whether a galaxy has experienced a quenching event in its past history. Using this criterion, we define $\qfrac(m_0)$ to be the fraction of galaxies in halos with $m_0$ that have $\qtime<t_0$.

We generally find that the quenched fractions in UM and TNG are quite similar: their values of $\qfrac(m_0)$ are within $10\%$ of each other at all mass. Moreover, the general trend of $\qfrac(m_0)$ is relatively simple, increasing smoothly and monotonically from $0.1$ at $m_0=11.5$ to $0.9$ for $m_0=13.5.$ However, when considering the full shape of $P(\qtime\vert m_0),$ the distributions in UM and TNG differ in their details. Generally speaking, $P(\qtime\vert m_0)$ is comparatively narrower in UM, so that TNG galaxies present a broader diversity of quenching times at fixed $m_0.$ Quenched galaxies in TNG show a smaller average value of $\qtime$ compared to UM for halos with $m_0 \lesssim 13$. This comparative trend reverses at higher mass, so that in massive halos with $m_0\gtrsim13$, galaxies in UM tend to quench earlier than in TNG.

\begin{figure}
\centering
\includegraphics[width=8cm]{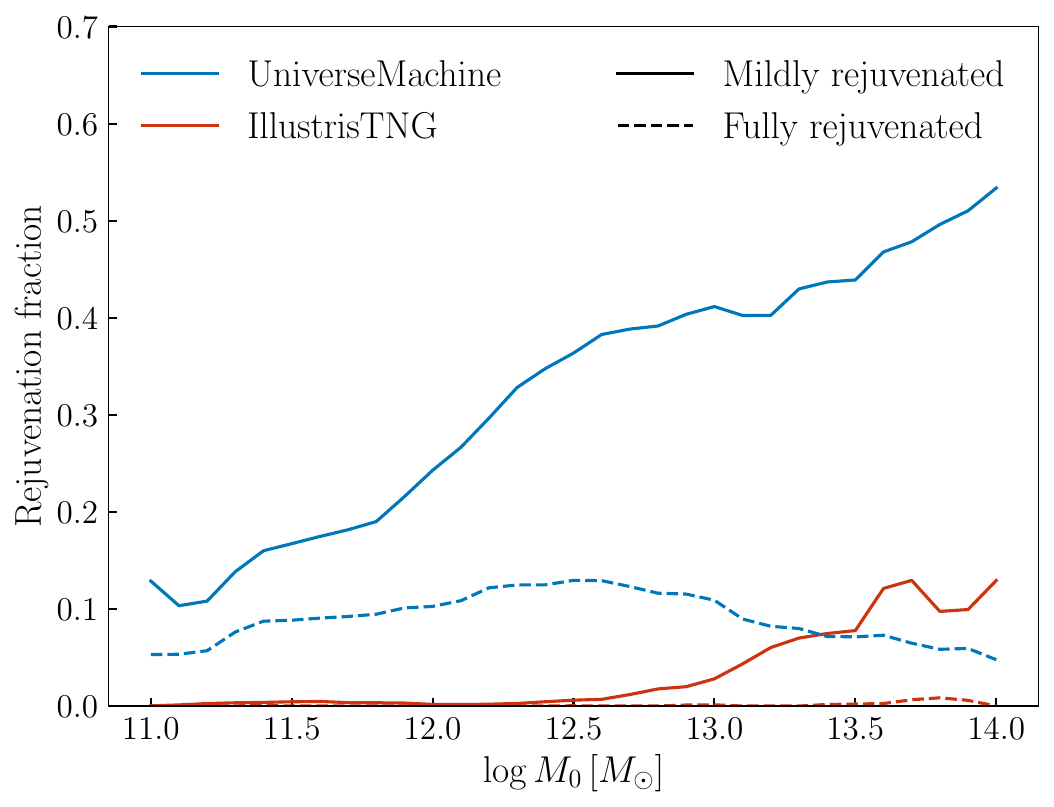}
\caption{{\bf Rejuvenation fraction}. The rejuvenation fraction of quenched galaxies as a function of the present day mass of host halos for UM (blue) and TNG (red). A large fraction of galaxies in UM experience mild rejuvenation ($q_{\rm drop}/q_{\rm rejuv}>10$) at some point in their history, with fractions as large as 50\% in massive halos. By contrast, very few galaxies rejuvenate in TNG, with only 10\% presenting some mild rejuvenation in massive halos. The fraction of fully rejuvenated galaxies ($q_{\rm drop}/q_{\rm rejuv}>100$) is roughly 5-10\% in UM, and practically negligible in TNG.
}
\label{fig:rejuv_frac}
\end{figure}

In Figure~\ref{fig:rejuv_frac}, we illustrate how the phenomenon of rejuvenation manifests in UM and TNG. To make a quantitative comparison between these two simulations, again we must define a specific criterion to designate when a quenched galaxy has experienced a rejuvenation event in its past history. Broadly speaking, a rejuvenated galaxy is one that has previously quenched (as defined above) and that also has $q_{\rm drop}\ll q_{\rm rejuv}.$ In designating whether or not a particular galaxy has experienced a rejuvenation event, we additionally require that its parameters satisfy $q_{\rm dt}>0.1\,{\rm dex},$ as we find that this criterion ensures that the rejuvenation event as a whole is non-trivial (smaller values of $q_{\rm dt}$ correspond to quenching events so rapid as to have an immaterial influence on the assembly history of the galaxy). We define $\rejuvfrac(m_0)$ to be the fraction of quenched galaxies in halos with $m_0$ that pass the rejuvenation cut, using $q_{\rm drop}/q_{\rm rejuv}>100$ to define {\it fully rejuvenated} galaxies, and $q_{\rm drop}/q_{\rm rejuv}>10$ to define {\it mildly rejuvenated} galaxies\footnote{A fraction of low mass TNG halos suddenly present a constant SFR=0 at late times \citep[e.g. Figure 1 in][]{Walters2022}. For such halos, our clip at $\rm{sSFR}=-12$ creates an artificial constant floor of SFR at late times. The \dstar fitter introduces a rejuvenation event to reproduce this constant floor, but such rejuvenation is physically immaterial. To filter out these cases from our presentation of the rejuvenation fraction, we additionally require that the SFR of the simulation at the quenching time $\qtime$ is non-zero, which we find by visual inspection is an effective criterion.}.

We find quite different behavior in the rejuvenation fractions between \um and \tng. In UM, a large fraction of previously quenched galaxies experience at least a mild rejuvenation event, especially for galaxies in massive halos, whereas even mild rejuvenation is generally rare in TNG. For quenched galaxies in massive halos, $\rejuvfrac$ reaches 50\% in UM when considering mild rejuvenation, whereas $\rejuvfrac\approx10\%$ for massive galaxies in TNG. When considering full rejuvenation, $\rejuvfrac\approx5-10\%$ for massive galaxies in UM, and is practically negligible in TNG.

\subsection{Halo Assembly Correlations}
\label{subsec:assembias}

Thus far within \S\ref{sec:interpretaish}, we have used the best-fitting \dstar parameters to compare the $m_0$-dependence of star formation history in \tng to \um, focusing on the main sequence in \S\ref{subsec:ms_comparison}, and on quenching in \S\ref{subsec:quenching_comparison}. In this section, we consider how SFH depends upon halo assembly history at fixed $m_0.$ We quantify this dependence in terms of $t^{\rm halo}_{\rm 50\%},$ the time at which the halo mass first exceeds half of its present-day value. Since the statistical distribution of $t^{\rm halo}_{\rm 50\%}$ itself depends upon $m_0$ (as shown in Fig.~\ref{fig:formation_times}), then the quantity we will use to study halo assembly correlations throughout this section is $\phalf,$ defined as $$\phalf\equiv P(<t^{\rm halo}_{\rm 50\%}\vert m_0),$$ the CDF of $t^{\rm halo}_{\rm 50\%}$ at fixed $m_0.$\footnote{We calculate $\phalf$ using the bin-free algorithm implemented in the \href{https://halotools.readthedocs.io/en/latest/api/halotools.utils.sliding\_conditional\_percentile.html}{sliding\_conditional\_percentile} function in {\tt halotools.utils} \citep{hearin_etal17_halotools}.} Thus halos with smaller values of $\phalf$ assemble their mass earlier relative to halos with larger values of $\phalf$ of the same $m_0.$ 

\begin{figure}
\centering
\includegraphics[width=8cm]{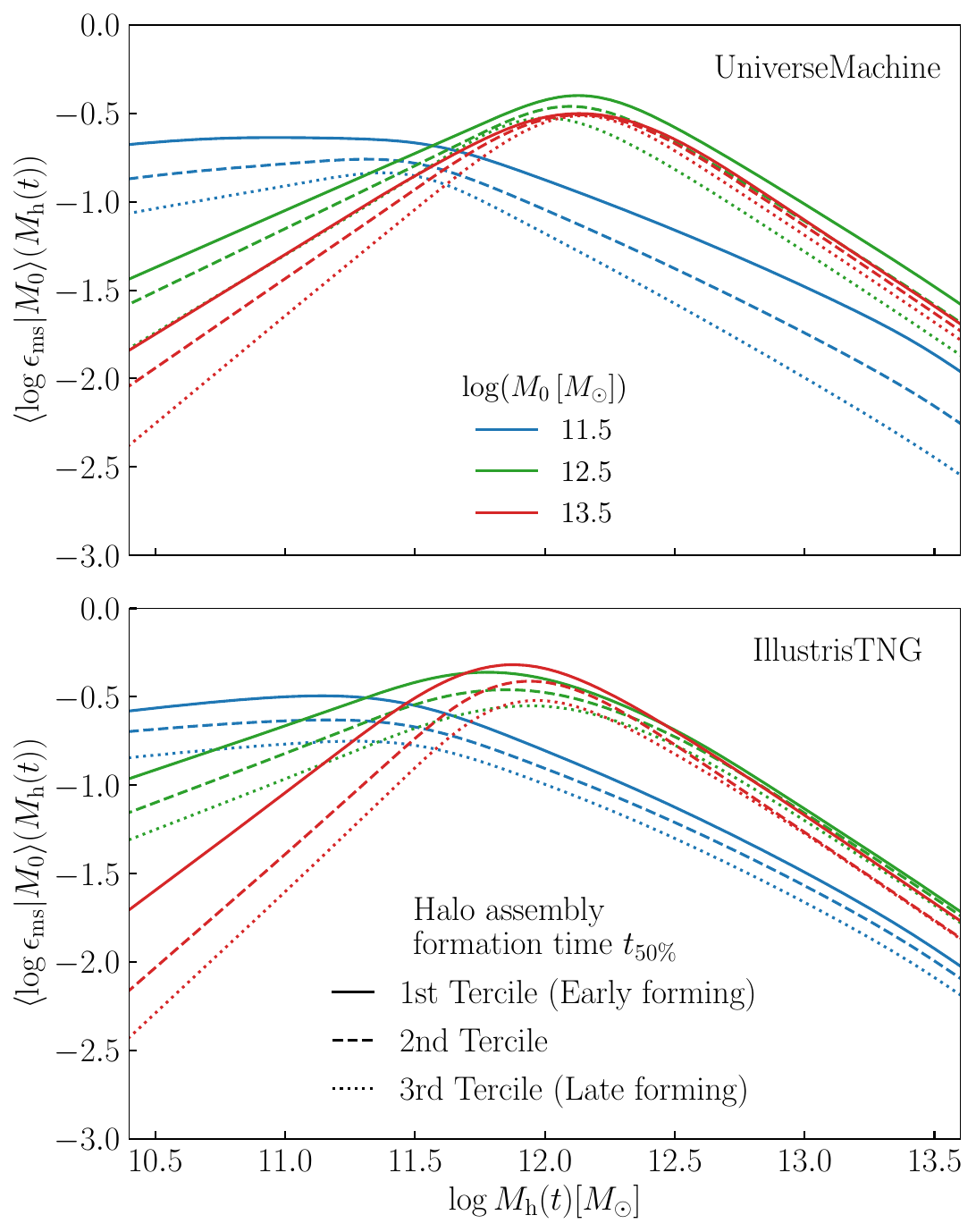}
\caption{{\bf Assembly bias of main sequence efficiency.} Same as Figure~\ref{fig:efficiency_results}, but for galaxies in halos of different mass and formation time. Early-forming halos tend to host galaxies with more efficient star formation. As a result, early-forming halos host more massive galaxies relative to late-forming halos with the same final mass. See the text for details.
}
\label{fig:assembly_bias_efficiency}
\end{figure}

\begin{figure}
\centering
\includegraphics[width=8cm]{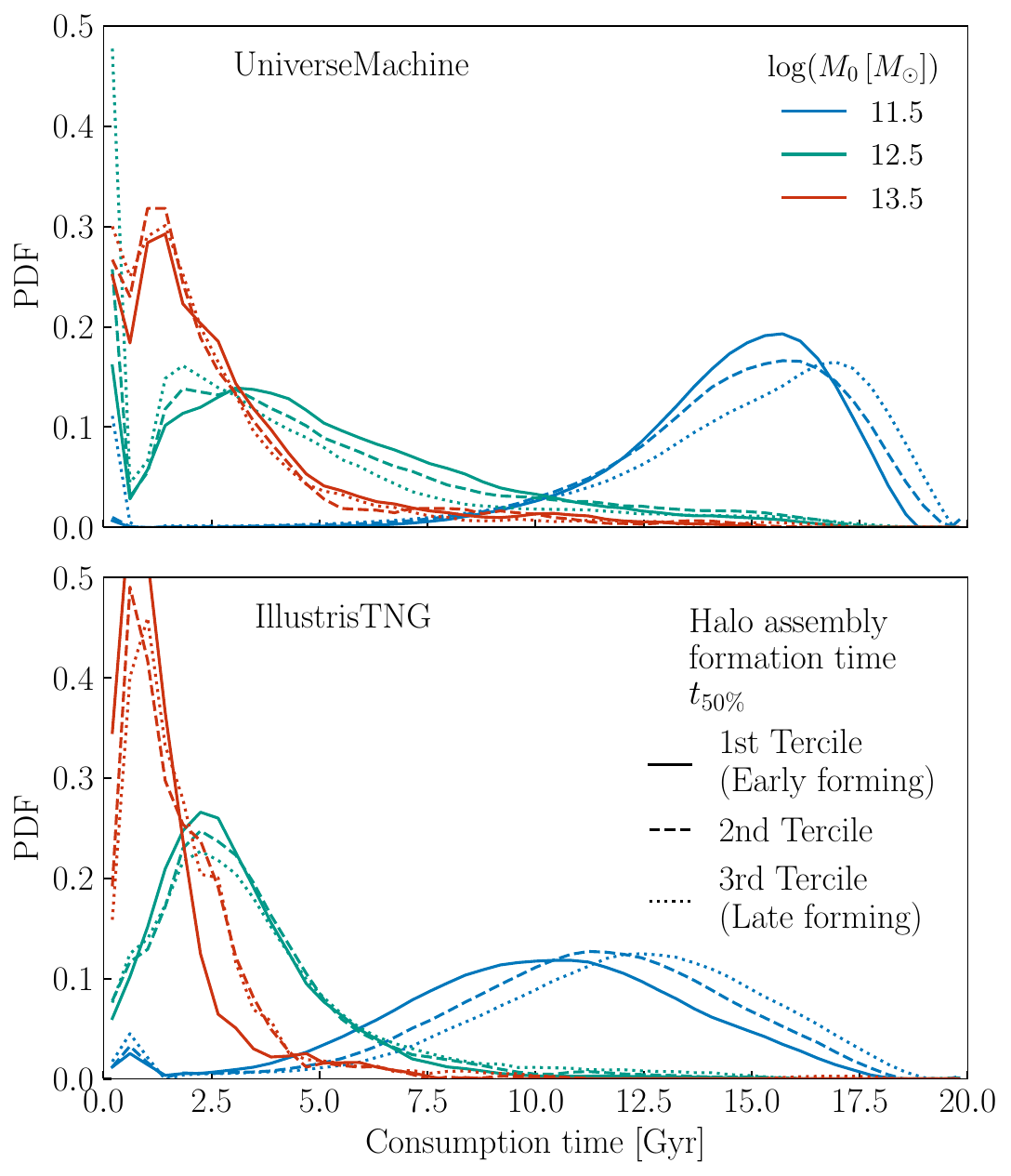}
\caption{{\bf Assembly bias of gas consumption timescales.} Same as Figure~\ref{fig:depletion_times}, but for galaxies in halos of different mass and formation time. Both UM and TNG present $\taucons$ distributions with similar halo assembly correlations. For the case of lower-mass halos, galaxies tend to have shorter gas consumption timescales if they formed earlier. For massive halos, we find no materially significant correlation between $\taucons$ and halo formation time.
}
\label{fig:assembly_bias_depletion_times}
\end{figure}

In Figure~\ref{fig:assembly_bias_efficiency}, we show how main sequence efficiency exhibits a {\it joint} dependence upon $m_0$ and halo assembly history, showing results for \um in the top panel, and for \tng in the bottom panel. As described in the legend, the $m_0$-dependence of $\mseff$ is encoded with curves of different colors, and the $\phalf$-dependence is encoded with curves with different line styles. Thus within each panel, comparing different curves of the same color illustrates how $\mseff$ varies with assembly history for halos of the same present-day halo mass.

For both simulations, we see that the solid curves lie above the dashed, which in turn lie above the dotted, indicating a clear dependence of star formation efficiency upon halo assembly. The sign of this trend is such that earlier-forming halos convert a significantly larger fraction of their accreted gas into stars relative to later-forming halos of the same present-day mass. In both simulations, this trend is especially strong for galaxies in lower-mass halos, and weakens in higher-mass halos. As a consequence of this phenomenon, we find that earlier-forming halos host more massive galaxies relative to later-forming halos of the same $m_0.$ For example, for galaxies in UM, the present-day average stellar mass of galaxies in halos with $m_0 = 11.5$ decreases as $\phalf$ increases, ranging from $m_\star = 9.5$ in the earliest-forming halos, to $9.2$ in the latest-forming halos; galaxies in TNG exhibit the same trend with comparable magnitude.

The joint dependence of gas consumption timescales upon $m_0$ and halo assembly is plotted in Figure~\ref{fig:assembly_bias_depletion_times}, which shows $P(\taucons)$ for UM in the top panel, and $P(\taucons)$ for TNG in the bottom panel. The interpretation of the color coding and line styles is the same as in Fig.~\ref{fig:assembly_bias_efficiency}, so that comparing different curves of the same color in Fig.~\ref{fig:assembly_bias_depletion_times} illustrates how the statistical distribution of $\taucons$ varies with the assembly history of halos of the same $m_0.$ For galaxies in massive halos in either simulation, $\taucons$ exhibits a weak-to-negligible correlation with $\phalf:$ the gas consumption timescale in massive galaxies is uniformly short, regardless of halo assembly history. The situation is different in lower-mass halos, where we find in both simulations that galaxies in later-forming halos have longer gas consumption timescales.

\begin{figure}
\centering
\includegraphics[width=8cm]{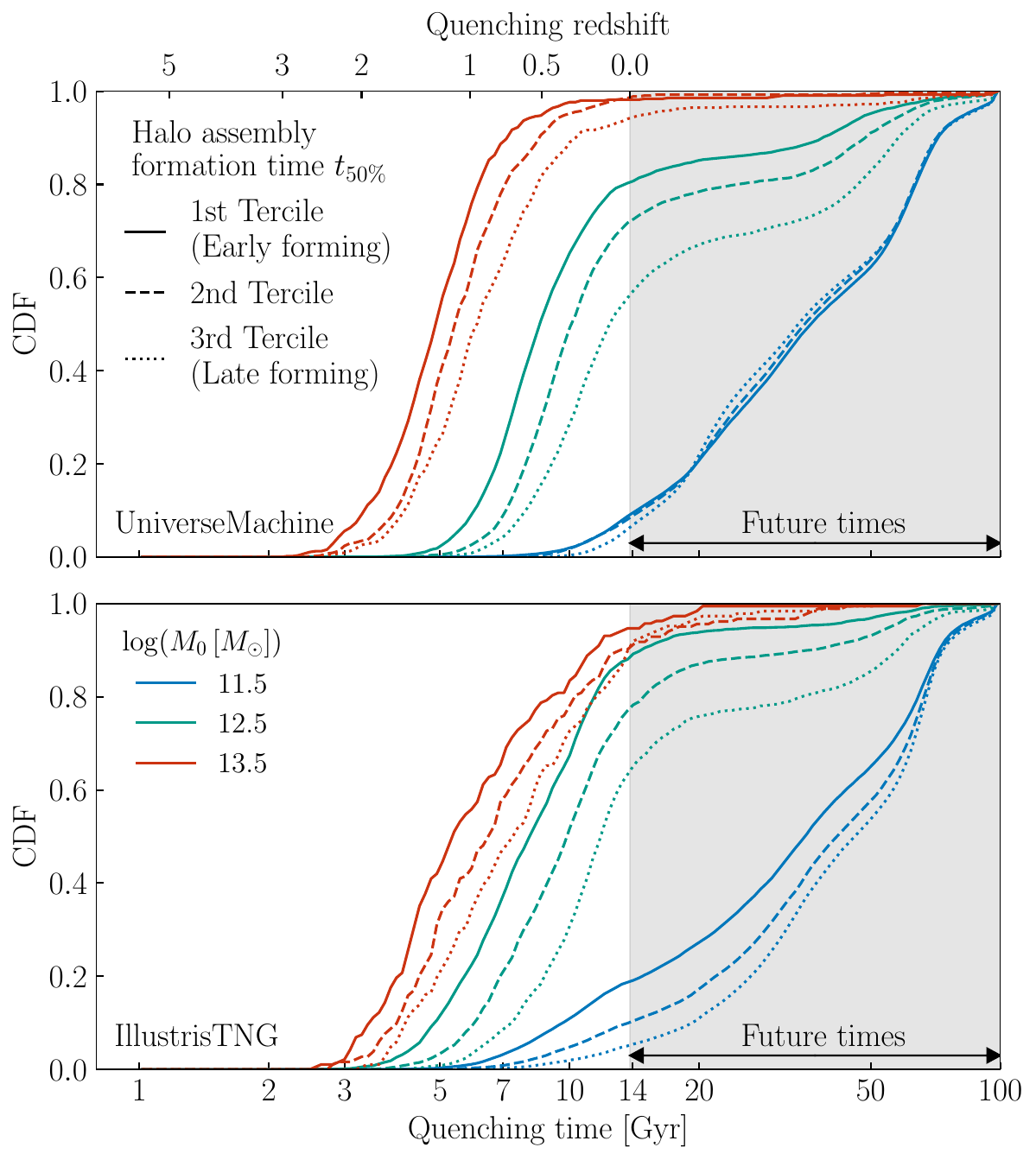}
\caption{{\bf Assembly bias of quenching times.} Same as Figure~\ref{fig:quenching_times}, but for galaxies in halos of different mass and formation time. In both TNG and UM, galaxies in earlier-forming halos are more likely to quench at earlier times relative to later-forming halos of the same final mass.
}
\label{fig:assembly_bias_quenching_times}
\end{figure}

Finally, in Figure~\ref{fig:assembly_bias_quenching_times} we turn attention to how the cumulative distribution of quenching times, $P(\qtime),$ exhibits a joint dependence upon $m_0$ and halo assembly. Again we use the same color-coding and line styles as in Figs.~\ref{fig:assembly_bias_efficiency} \& \ref{fig:assembly_bias_depletion_times}, so that comparing different curves of the same color in Fig.~\ref{fig:assembly_bias_quenching_times} illustrates how the statistical distribution of quenching times varies with the assembly history of halos of the same $m_0.$ In both simulations, and for halos of all mass $m_0\gtrsim12,$ Fig.~\ref{fig:assembly_bias_quenching_times} shows  that earlier-forming halos host galaxies that are significantly more likely to experience a quenching event. Additionally, when considering quenched galaxies residing in halos of the same mass, $P(\qtime)$ is shifted to earlier times for earlier-forming halos, and conversely for later-forming halos. Thus in UM and TNG alike, both the quenched fraction $\qfrac(\qtime < t_0)$, and the average quenching time, correlate strongly with $\phalf.$ The magnitude of this effect is especially pronounced for galaxies in halos with $m_0=12.5:$ in UM, $\qfrac=0.84$ for galaxies in early-forming halos, while $\qfrac=0.61$ in later-forming halos ($0.89$ and $0.63$ in TNG).

\section{Discussion \& Future Work}
\label{sec:discussion}

We have presented a new parametric model for the in-situ star formation history of galaxies, \dstar. As discussed in \S\ref{sec:intro}, there are a wide variety of parametric forms for SFH that have been in the literature for years; a distinguishing feature of our parametrization is that it is explicitly defined in terms of basic features of galaxy formation physics. Built into the formulation of the \dstar model are the assumptions that the accretion rate of gas, $\dot{M}_{\rm g},$ is proportional to the accretion rate of the dark matter halo, $\dot{M}_{\rm halo},$ and that main sequence galaxies transform accreted gas into stars with efficiency, $\mseff,$ over a gas consumption timescale, $\taudep.$ Furthermore, there is parametric freedom to capture the phenomenon of quenching, which can be either rapid or gradual, as well as freedom incorporating the possibility that a quenched galaxy may subsequently experience rejuvenated star formation.

\subsection{\dstar flexibility}

In order to validate the flexibility of our model, we have fitted the \dstar parameters to hundreds of thousands of SFHs taken from \um and \tng. The physical assumptions underlying these two simulations are quite distinct: TNG is a hydrodynamical simulation with sub-grid prescriptions for baryonic feedback, while UM empirically maps SFR to subhalos at each snapshot of a gravity-only simulation. In both cases, the SFHs of individual galaxies are determined from the merger trees in post-processing, and so are {\it emergent}, and thus do not admit an exact parametric description. This guarantees that \dstar can at best be only an approximate description of the SFHs in these simulations. We have used TNG and UM to demonstrate that \dstar is formulated on sufficiently sound physical principles, and is sufficiently flexible, to describe the stellar mass histories (SMHs) of both simulations in an unbiased fashion to within 0.1 dex across most of cosmic time. 

Recent works studying the influence of SFH on galaxy SEDs have highlighted the shortcomings of traditional parametric models. In \citet{carnall_etal19} and \citet{leja_etal19}, it was shown that conventional forms such as a lognormal or a (delayed) exponentially declining function can lead to significant biases in measurements of basic scaling relations such as the star-forming sequence, and that these biases can be considerably reduced by instead using more flexible, piecewise-defined SFH models. These companion papers demonstrated that the functional forms of traditional parametric models rule out in advance a significant fraction of SFH shapes, and called attention to the risk of unintentionally imposing prior assumptions about the true distribution of galaxy assembly histories. Broadly speaking, SFH models such as piecewise-defined formulations have been designed to address these concerns by being flexible enough to describe arbitrarily complex SFH shapes, and by adopting priors that are as uninformative as possible.

We have formulated the \dstar model to address these same concerns with a complementary approach. First, we have eschewed the goal of capturing arbitrarily complex SFHs, and instead sought to identify the {\it minimum} parametric flexibility that is required to accurately capture the SFHs in \um and \tng. Second, we have embraced the ill-conditioned nature of the SFH inference problem \citep{Ocvirk2006}, and deliberately adopted physics-informed priors on the shape of galaxy assembly history. The key to our approach is that \dstar is formulated in terms of elementary physical ingredients that we expect to pertain to all galaxies, or at least to the vast majority of the population. Thus by approximating galaxy SFHs with \dstar, we intentionally impose basic features of galaxy formation physics onto the interpretation of the measurement. We return to this point below when we discuss \DstarPop, a hierarchical Bayesian model characterizing populations of \dstar SFHs.

In order to establish that the \dstar parameterization contains sufficient flexibility, we have directly fitted the SFHs of galaxies in \um and \tng. These two simulations have been shown to exhibit good agreement with a wide range of observations, and the assumptions underlying these two models are moreover radically different from one another, and so the validation exercises presented here are highly nontrivial. However, as pointed out in \citet{leja_etal19}, even if the direct fits to ${\rm SFH}(t)$ are acceptable, the model could still perform rather poorly when using it to interpret observed SEDs, because for galaxies with continued ongoing star formation, the early-time SFH shape has a rather minimal influence on the late-time SED. Thus when a range of SFH shapes conspire to produce similar SEDs, a successful model should effectively up-weight the physically realistic SFH, and down-weight the unrealistic alternatives. This can be achieved with piecewise-defined models by tuning the parameters that specify the otherwise uninformative prior \citep[as in, e.g.,][]{iyer_etal19,leja_etal19_3dhst}, whereas in our case this relative weighting will be accomplished through a combination of the physically-motivated priors supplied by \DstarPop, as well as through the hard-wiring of galaxy formation physics into the functional forms of \dstar. In order to assess the level of success that is achievable with \dstar in such SED-fitting applications, it will be necessary to carry out an analogous program as presented in \citet{lower_etal20}, in which the residual errors in SFH approximations are fully propagated through to SED fits. We aim to conduct this exercise in future work in which we subject \dstar to additional validation data based on semi-analytic models and alternative hydrodynamical simulations.

\subsection{\dstar physical interpretation}

As an additional advantage of formulating our parametrization in terms of basic features of galaxy formation, \dstar can be used to provide a physical comparison between simulations that are founded upon disparate assumptions. In \S\ref{sec:interpretaish}, we showed how \dstar reveals striking similarities between the UM and TNG models. Both simulations are characterized by very similar distributions of baryon conversion efficiency, $\mseff,$ gas consumption timescale, $\taucons,$ and quenching time, $\qtime;$ moreover, the distributions of these quantities in UM and TNG each share the same $\Mhalo$-dependence in reasonably quantitative detail. In both simulations, $\mseff(\Mhalo)$ is well described by a broadly similar shape, with a monotonic falloff in efficiency on either side of a peak at $\mcrit\approx12$. The $m_0$-dependence of $P(\taucons)$ is also very similar between UM and TNG, being sharply peaked around $\taucons\approx1-2$ Gyr for galaxies in massive halos, and broadly distributed at $\taucons>10$ Gyr for $m_0\approx11.5.$ Finally, for galaxies in both simulations, $P(\qtime)$ peaks at earlier times in more massive halos, and relatively few galaxies in halos with $m_0\lesssim12$ experience a major and sudden quenching event. We posit that {\it any} successful model of galaxy formation should possess these same basic trends in $\mseff,$ $\taucons,$ and $\qtime,$ and we predict that these trends will be directly revealed in observational data when \dstar is used to fit the SEDs of large samples of galaxies (see below for further discussion of such applications). 

\subsection{Galaxy assembly bias}

Our analysis in \S\ref{sec:interpretaish} also reveals how $\mseff,$ $\taucons,$ and $\qtime$ correlate with halo assembly history in these two simulations. When considering a population of galaxies that reside in halos of the same mass, statistical correlations between galaxy properties and halo assembly history are generally referred to as {\it galaxy assembly bias}. This term is typically defined in terms of the occupation statistics of dark matter halos: using $P(N_{\rm gal}\vert\Mhalo)$ to denote the statistical distribution of the number of galaxies (of a given type) residing in halos of mass $\Mhalo,$ it is said that the occupation statistics exhibit galaxy assembly bias if $$P(N_{\rm gal}\vert\Mhalo)\neq P(N_{\rm gal}\vert\Mhalo,\tform),$$ where $\tform$ is some marker of halo assembly history \citep[see, e.g.,][]{zentner_etal14,wechsler_tinker18}.\footnote{We note that galaxy assembly bias is a special case of {\it secondary galaxy bias}, in which $P(N_{\rm gal}\vert\Mhalo)\neq P(N_{\rm gal}\vert\Mhalo,X_{\rm halo}),$ where $X_{\rm halo}$ is any halo property that is correlated with the cosmic density field, but need not be directly related to halo assembly history \citep[see][for further details]{salcedo_etal18,mao_etal18}.} This is the same definition adopted in recent studies of the occupation statistics of galaxies in \tng, and it has by now been firmly established that galaxy assembly bias of this form is rather strong in the TNG model \citep{artale_zehavi_etal18_tng_eagle_assembias,bose_etal19_tng_hod_assembias,hadzhiyska_etal20_basic_hods,hadzhiyska_etal21_tng_assembias,hadzhiyska_etal21_tng_elgs,montero_dorta_etal21_tng_assembias,yuan_etal22_tng_lrgs}.

Because \dstar is parametrized in terms of basic features of galaxy formation, our model enables us to study a novel form of galaxy assembly bias that is defined in terms of how individual physical ingredients may be correlated with halo assembly at fixed present-day mass. In both simulations, $\mseff$ is larger in earlier-forming halos relative to later-forming halos with the same $m_0,$ and for halos of mass $m_0\gtrsim12,$ earlier-forming halos host galaxies that are significantly more likely to experience a quenching event. Furthermore, we find that in both simulations $\taucons$ correlates strongly with $\tform$, especially for galaxies in lower-mass halos. In our analysis of the results presented in Figure~\ref{fig:formation_times}, we have shown how the statistical connections between $\mseff,$ $\taucons,$ and halo properties imprint a signature upon the correlation between galaxy and halo formation time. In future work, we will further study how halo assembly correlations in the Diffstar parameters of galaxy populations are closely connected to conventional halo occupation-based notions of galaxy assembly bias. Additionally, we aim to use our model to infer the true strength of these halo assembly correlations from cosmological measurements of large-scale structure (see discussion below of \DstarPop).

\subsection{Improving \dstar}

Although this paper has presented progress on the parametric modeling of galaxy SFH, our work on \dstar is still ongoing, and there are several aspects of our model that will benefit from further improvement. 

\subsubsection{High redshift}
The fidelity with which \dstar recovers the SFH of simulated galaxies degrades at high redshift, particularly for lower-mass halos (see Figs.~\ref{fig:average_Mstar}$-$\ref{fig:residuals_sfh}). There are several possible origins for this shortcoming. First, the formulation of the \dstar model could simply be insufficiently flexible to accurately capture the full diversity of star formation at high redshift. Alternatively, since \dstar is defined in terms of the \dmah parametrization of dark matter halo assembly \citep{hearin_etal21_diffmah}, then if \dmah were insufficiently flexible at high redshift, \dstar would inherit this limitation. In either case, the remedy would be the addition of extra parametric freedom at early times. However, since lower-mass halos/galaxies in both \um and \tng are only resolved with tens-to-hundreds of particles at high redshift, it would be necessary to proceed with care to protect against over-fitting, and so such an extension would best be carried out in concert with a dedicated resolution study. Of course, the observational data used to constrain the SFHs of low-mass halos at high redshift in these two simulations is only weakly constraining, and so another possibility is that updates to \um and \tng based on future measurements will ameliorate the discrepancy in this regime. We have explored several alternative formulations of our model for $\mseff(\Mhalo)$ that include additional flexibility at low mass designed to address this issue, but for present purposes we decided to relegate a proper investigation to future work since our own shorter-term aims are to use \dstar to interpret large-scale structure measurements of $\Mstar\gtrsim10^{10}\msun$ galaxies at $z\lesssim2.$ However, the resolution to this issue may have important consequences for the physics of dwarf galaxy formation and the stellar-to-halo-mass relation in low-mass halos, and so the effort to better understand \dstar in this regime is scientifically well-motivated.

\subsubsection{Residual SFH burstiness}

The \dstar model primarily captures changes in star formation over timescales $\Delta t\gtrsim1$ Gyr, but not short-term transient fluctuations, i.e., burstiness. As shown in Figure~\ref{fig:residual_SFR_variance}, when focusing on timescales $\Delta t\approx100$ Myr, there exists a typical scatter of $\sim0.3$ dex in the residuals of the \dstar fits. As pointed out in \citet{carnall_etal19}, the lack of a treatment of bursts is a significant contributor to the biases of traditional parametric models, and so incorporating bursts into \dstar is a focal point of our ongoing work. Our approach is closely related to recent modeling efforts studying the power spectrum density of simulated SFHs \citep[e.g.,][]{iyer_etal20,tacchella2020_StochasticModellingstarformation}. As shown in \citet{chaves_montero_hearin_2021_sbu2}, when the SFHs in \um and \tng are fitted with an unbiased model for longer-timescale fluctuations, the power spectrum of the {\em residuals} on shorter timescales is well approximated by a simple power law. Motivated by these findings, our approach is to implement the bursts as separate (differentiable) episodes that are layered on top of the smooth history captured by \dstar, and to statistically distribute these episodes according to a parametrized power spectrum. We will present results based on a bursty version of \dstar in a follow-up paper to the present work.

\subsubsection{Halo boundary definition}

In the validation exercises presented here, we have not explored the impact of the choice of the halo boundary on the inferred star formation histories. However, low-redshift halos commonly accrete mass in their outskirts while experiencing little-to-no growth of the physical density at the halo center \citep{diemer_etal13,more_etal15_splashback_radius_is_physical}; moreover, the relative influence of substructure on estimates of $\dmqdt{\Mhalo}$ also exhibits significant dependence upon how the halo boundary is defined \citep{diemer_kravtsov_2014,mansfield_etal17_splashback_shells}. These results indicate that SFHs inferred via \dstar may be more realistic when using physically-motivated halo boundary definitions such as those based on particle dynamics \citep[e.g.,][]{diemer17_sparta1,diemer21_dynamics_based_halo_profiles}. A proper study of this important issue would require a hydrodynamical simulation with merger trees based on homogeneously processed catalogs of halos/galaxies that have been identified with multiple choices for the halo boundary; we leave this investigation as a task for future work when such data products become publicly available.

\subsubsection{Halo gas mass}

As discussed in Appendix~\ref{appendix:sams}, the particular shape of the gas consumption function is an emergent feature of gas being consumed in accord with the baryon conversion efficiency of main sequence galaxies. For example, a parcel of gas that is accreted by the halo and subsequently consumed at a constant rate will produce a power-law shaped consumption function. Alternatively, if the baryon conversion efficiency increases with time (as is the case for halos that remain below their critical mass), then the consumption function takes a bell-like shape like the one shown in Figure~\ref{fig:depletion}. Although $\mseff$ and $\taucons$ are implemented separately in Diffstar, there is at some level a degeneracy between the associated parameters when predicting SFH alone. It is also possible that alternative parametrizations of these functions could give similarly accurate predictions of these SFHs. This degeneracy can be broken by jointly fitting the star formation histories together with the halo gas mass histories, as \dstar makes a specific prediction for $\Mgas(t)$ that is currently going untested. Such predictions will be especially relevant when we use Diffstar in future work to predict the Sunyaev-Zeldovich effect, as well as to predict the gas content of the circumgalactic medium.

\subsection{DiffstarPop: a simulation-based model of the galaxy–halo connection}
As the principal aim of our future work, we are using \dstar as the foundation of a new, simulation-based model of the galaxy--halo connection, \DstarPop. The work presented in \citet{hearin_etal21_diffmah} to develop the \dmah model provides a blueprint for this effort. The first stage of developing \dmah was the calibration of a fitting function for $\Mhalo(t),$ the mass assembly history of individual dark matter halos across time; this calibration required establishing that the \dmah fitting function has sufficient flexibility to accurately capture $\Mhalo(t)$ for halos simulated in both gravity-only and hydrodynamical simulations, using a directly analogous metric to Figure~\ref{fig:residuals_sfh} of the present work. In the second stage of developing \dmah, we studied the statistical distribution of the best-fitting parameters of the fitting function, and built a parametrized, analytical model for the PDF. The resulting hierarchical model, DiffmahPop, has the capability to generate cosmologically representative populations of dark matter halo assembly histories. Our ongoing work to develop \DstarPop mirrors this two-stage process; the aim of our effort is to produce a hierarchical model that can generate cosmologically representative populations of galaxy star formation histories, including the capability to map \dstar SFHs  onto the individual merger trees of a simulated population of dark matter halos.

We anticipate two classes of applications of this hierarchical model. First, when fitting the SEDs of individual galaxies, \DstarPop will supply the {\it priors} on the \dstar parameters, thereby allowing us to derive constraints on the physical properties of observed galaxies under ``UM-like" or ``TNG-like" assumptions about the physics of galaxy formation. Since \dmah is an ingredient of \dstar, this class of applications has the exciting potential to use measurements of the SED of a galaxy to infer the mass assembly history of its parent dark matter halo in a fully Bayesian fashion \citep[see][for a related effort based instead on machine learning]{Eisert2022}. Second, \DstarPop together with DSPS \citep{hearin_etal22_dsps} will allow us to populate large-volume cosmological simulations with proposed populations of galaxies, and to use observations of two-point functions such as galaxy clustering and lensing to constrain the parameters of the hierarchically-formulated model of the galaxy--halo connection. We note, however, that since the full diversity of observed galaxies in color-color space cannot be accounted for by SFH variability alone \citep[see, e.g.,][]{chaves_montero_hearin_2020_sbu1}, then achieving this predictive power will require additional modeling of how SPS ingredients such as dust and metallicity vary across the galaxy population \citep[see][for recent progress along these lines in the case of galaxy attenuation]{Nagaraj_etal22,lower_etal22,hahn_etal22_dust_model}. When coupled with the cosmological evidence modeling technique \citep{lange_etal19,lange_etal22}, together with methods for making differentiable predictions for large-scale structure \citep{hearin_etal21_shamnet}, this second class of applications opens up the possibility to derive cosmological constraints from multi-wavelength observations of multiple tracer populations across redshift in a joint analysis. 

A novel feature of this hierarchical modeling program is that it enables one to fit observational measurements of both individual galaxies, as well as galaxy populations, with the exact same underlying model. In likelihood analyses of conventional semi-analytic models (SAMs), the SEDs of some observed galaxy sample(s) are first fit with an SPS model that makes one set of SFH assumptions, and then summary statistics measured in the resulting  catalog(s) of galaxies (e.g., the stellar mass function, SMF) are used to derive constraining data for the SAM parameters, even though the SAM implicitly makes a distinct set of assumptions about galaxy SFH.  In the hierarchical modeling program outlined above, summary statistics such as the SMF and galaxy clustering will instead be self-consistently interpreted with the exact same SFH and SPS models, simplifying the direct use of posteriors from one analysis as the priors in another analysis. The development of the \dstar model presented in this work is an important milestone towards achieving the goals of this program.

\section{Conclusions}
\label{sec:conclusions}

We conclude by summarizing our primary results:

\ben
\item We have introduced \dstar, a fully parametric model for the in-situ star formation history (SFH) of individual galaxies. The \dstar parameters are physically interpretable in terms of basic features of galaxy formation physics. We have shown that our model is sufficiently flexible to describe the stellar mass histories of central galaxies in \um (UM) and \tng (TNG) with an accuracy of $\sim0.1$ dex across most of cosmic time.
\item We have used \dstar to physically interpret the SFHs in UM and TNG. Regarding trends between key physical properties and halo mass, we find:
\bit
\item Star formation efficiency in \dstar is quantified in terms of $\mseff,$ defined as the fraction of accreted gas that eventually turns into stars. In UM, star formation is less efficient relative to TNG, particularly in lower-mass halos. 
\item The timescale for gas consumption, $\taudep,$ is longer in UM relative to TNG, with both simulations presenting a strong dependence upon halo mass. In both simulations, the distribution $P(\taudep)$ is tightly peaked around $\taucons\approx1-2$ Gyr for massive halos, and broadly distributed at $\taucons\gtrsim10$ Gyr in lower-mass halos.
\item For galaxies residing in halos of all mass, star formation in UM is burstier relative to TNG.
\item The distribution of quenching times, $P(\qtime),$ is also similar in UM and TNG. In both simulations, $P(\qtime)$ peaks at earlier times ($z\approx1-2$) for massive galaxies, while very few centrals in low-mass halos experience a major quenching event. However, TNG presents a broader diversity of quenching times relative to UM, particularly in massive halos.
\item Rejuvenated star formation of previously-quenched centrals is ubiquitous in UM, particularly in massive galaxies, whereas quenched centrals in TNG rarely rejuvenate.
\eit
\item We have additionally studied how the best-fitting \dstar parameters correlate with halo assembly time, $\tform,$ at fixed present-day halo mass, $m_0.$ In this way, our model allows us to dissect the phenomenon of galaxy assembly bias in terms of basic physical processes. We find:
\bit
\item In both simulations, $\mseff$ is larger in earlier-forming halos relative to later-forming halos with the same $m_0.$ This correlation between star formation efficiency and $\tform$ manifests in earlier-forming halos hosting more massive galaxies relative to later-forming halos of the same $m_0.$
\item In both simulations, $\taucons$ correlates strongly with $\tform$ for galaxies in halos with $m_0=11.5;$ this correlation weakens with increasing halo mass, and vanishes for $m_0\gtrsim13.5.$ 
\item In both simulations, and for halos of all mass $m_0\gtrsim12,$ earlier-forming halos host galaxies that are significantly more likely to experience a quenching event. Additionally, when considering quenched galaxies residing in halos of the same mass, $P(\qtime)$ shifts to earlier times for earlier-forming halos, and shifts to later times for later-forming halos.
\eit
\item A JAX-based software implementation of our model, \href{https://github.com/ArgonneCPAC/diffstar}{\dstarcode}, is publicly available on GitHub. The \dstarcode library can be installed with conda-forge or pip.
\een

\section*{Acknowledgements}
Argonne National Laboratory's work was supported by the U.S. Department of Energy, Office of High Energy Physics. Argonne, a U.S. Department of Energy Office of Science Laboratory, is operated by UChicago Argonne LLC under contract no. DE-AC02-06CH11357. We gratefully acknowledge the computing resources provided on Bebop and Swing, high-performance computing clusters operated by the Laboratory Computing Resource Center at Argonne National Laboratory. This material is based upon work supported by Laboratory Directed Research and Development (LDRD) funding from Argonne National Laboratory, provided by the Director, Office of Science,  of the U.S. Department of Energy under Contract No. DE-AC02-06CH11357. This work was performed in part at the Aspen Center for Physics, which is supported by National Science Foundation grant PHY-1607611. JCM acknowledges partial support from the Spanish Ministry of Science, Innovation and Universities (MCIU/AEI/FEDER, UE) through the grant PGC2018-097585-B-C22.

We thank the developers of {\tt NumPy} \citep{numpy_ndarray}, {\tt SciPy} \citep{scipy}, Jupyter \citep{jupyter}, IPython \citep{ipython}, JAX \citep{jax2018github}, conda-forge \citep{conda_forge_community_2015_4774216}, and Matplotlib \citep{matplotlib} for their extremely useful free software. We thank Peter Behroozi and Benedikt Diemer for useful discussions. While writing this paper we made extensive use of the Astrophysics Data Service (ADS) and {\tt arXiv} preprint repository. The authors gratefully acknowledge the Gauss Centre for Supercomputing e.V. (www.gauss-centre.eu) and the Partnership for Advanced Supercomputing in Europe (PRACE, www.prace-ri.eu) for funding the MultiDark simulation project by providing computing time on the GCS Supercomputer SuperMUC at Leibniz Supercomputing Centre (LRZ, www.lrz.de). The Bolshoi simulations have been performed within the Bolshoi project of the University of California High-Performance AstroComputing Center (UC-HiPACC) and were run at the NASA Ames Research Center.

\section*{Data Availability}
\begin{itemize}
    \item The code for this work is publicly available at: \url{https://github.com/ArgonneCPAC/diffstar}
    \item The data for this work is publicly available at: \url{https://portal.nersc.gov/project/hacc/aphearin/diffstar_data/}
\end{itemize}



\bibliographystyle{aasjournal}
\bibliography{bibliography}




\appendix
\counterwithin{figure}{section}

\renewcommand{\thefigure}{A\arabic{figure}}

\section{Diffstar Model Definition}
\label{appendix:model_def}

In this section, we provide a concise statement of each ingredient of our parametric model for the star formation history (SFH) of an individual galaxy. This appendix is intended to complement the lengthier and more pedagogical material in \S\ref{sec:model_formulation} by serving as a compact summary of the defining equations and assumptions of our model.

We assume that stars form from accreted baryonic mass, and that the gas accretion rate is proportional to the accretion rate of the dark matter halo, $$\dmqdt{\Mgas}=f_{\rm b}\cdot\dmqdt{\Mhalo}.$$  For a dark matter halo with present-day mass $\mzero,$ we use the Diffmah model to approximate the halo assembly history:$$\Mhalo(t) = \mzero(t/t_0)^{\alpha(t)},$$ where $\alpha(t)$ decreases smoothly and monotonically with time.

For a small parcel of gas, $\delta\Mgas,$ that accretes at some time, $t,$ the portion of this mass that turns into stars at some later time, $t'>t,$ is controlled by the {\em baryon conversion efficiency}, $\mseff(\Mh(t')),$ which is defined by the following proportionality:
\beq
\label{eq:bce_appendix}
\delta\Mstar(t')\propto\mseff(\Mh(t'))\times\delta\Mgas(t).\nonumber
\eeq
We model the halo mass dependence of the baryon conversion efficiency as
\beq
\label{eq:mseff_appendix}
\mseff(\Mh) = \epsilon_{\rm crit}\cdot(\Mh/\Mcrit)^{\beta(\Mh)},\nonumber
\eeq
where at low mass, $\mseff(\Mh)$ increases monotonically until reaching a peak value near $\Mcrit\approx10^{12}\Msun,$ and then decreases monotonically at higher mass (see Eqs.~\ref{eq:mseff}-\ref{eq:diffmah_slope}).

We assume the conversion of accreted gas into stellar mass is a gradual process that occurs over the {\it gas consumption timescale,} $\taucons.$ Thus the star formation rate of a ``main sequence" galaxy, $\sfrms,$ receives a contribution from all the previously accreted parcels of gas, $\Mgas(t'),$ for all times $t-\taucons\leq t'\leq t:$
\beq
\label{eq:mssfr_appendix}
\sfrtfracms=\mseff(\Mh(t)) \int_{0}^{t}\dtime'\dmqdtpfrac{\Mgas(t')}\cdot\fcons(t\vert t',\taucons)\nonumber.
\eeq
The function $\fcons(t\vert t',\taucons)$ defined in Eq.~\ref{eq:fdep} controls the gradual transformation of accreted gas into stars.

In \dstar, star formation rates can plummet below the main sequence value according to the quenching function, $\Qfunc(t),$ which acts as a multiplicative factor on the main sequence star formation rate:
\beq
\label{eq:quenching_appendix}
\sfrtfrac = \Qfunc(t)\times\sfrtfracms.\nonumber
\eeq
The quenching function is defined by Eq.~\ref{eq:qfunc} in terms of a logarithmic drop in SFR; quenching in \dstar can be either rapid or gradual, and the behavior of $\Qfunc(t)$ includes flexibility to account for the possibility of a rejuvenation event that returns the galaxy to the main sequence.

\section{Relationship to Semi-analytic Approaches}
\label{appendix:sams}

\begin{figure}
\centering
\includegraphics[width=8cm]{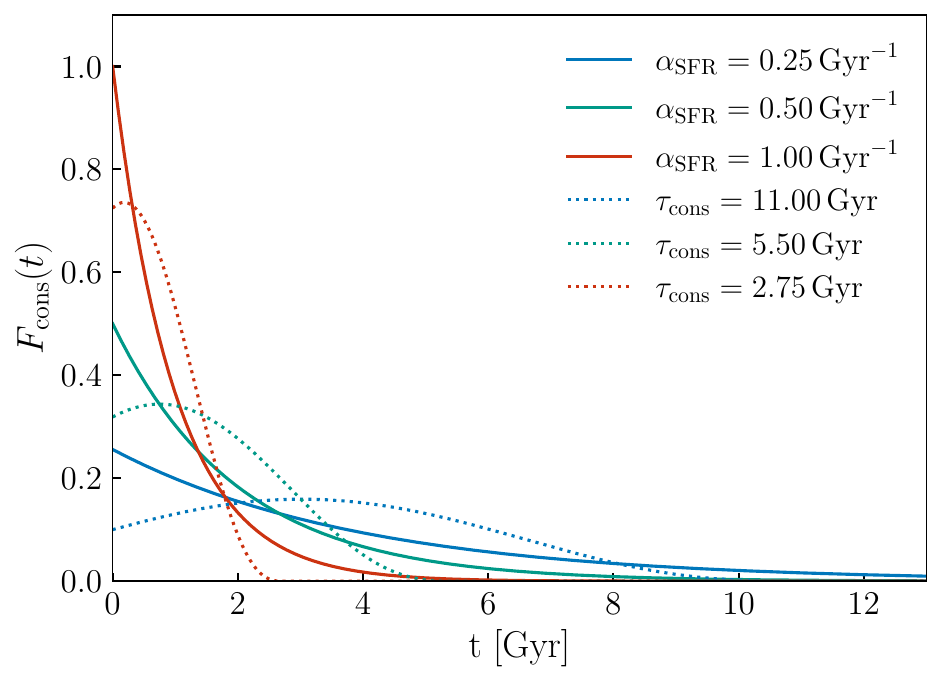}
\caption{{\bf Consumption functions in SAM-like models}. The solid lines show the consumption function that emerges in a one-parameter family of SAM-like models, with values of $\alpha_{\rm SFR}$ color-coded according to the legend. Each solid line is accompanied by a dotted line of the same color that shows the behavior of the $\fcons$ function parametrized in \dstar. The figure shows that the consumption functions assumed in this work are an approximation to the \textit{emergent} properties of more fine-grained SAMs.}
\label{fig:sam_dep}
\end{figure}

In this appendix, we discuss the mathematical formulation of some semi-analytic models of galaxies (SAMs) in comparison to \dstar, focusing in particular on the gas consumption timescale, $\taucons.$ In most SAMs, star formation history is only one among numerous time-evolving reservoirs of material associated with a galaxy (e.g., stars, cold and hot gas, metals, etc.). The time evolution of these quantities is determined via a solution to a system of ordinary differential equations (ODEs) that regulate the dynamic exchange of material between the collection of reservoirs. In particular, the ODE system of a SAM controls how a galaxy's reservoir of gas in the interstellar medium (ISM) fuels star formation over time, and so a time-lagged transformation of accreted gas into stars should be a generic, {\it emergent} prediction of SAMs. In \dstar, we have assumed a particular form for a one-parameter family of functions that serve as an approximation to this transformation, the {\it gas consumption function}, $\fcons(t\vert\taucons),$ defined in Eq.~\ref{eq:fdep} and illustrated in Fig.~\ref{fig:depletion}. In the remainder of this section, we construct a simplified ODE that we use as a toy SAM to understand how the Diffstar parameter $\taucons$ relates to the treatment of star formation in traditional semi-analytic modeling.

Motivated by the long-standing observation that the surface density of star formation and gas are proportional \citep{Schmidt59,Kennicutt89}, let us begin with the common assumption \citep[e.g.,][]{fu_etal13_sfr_disks_Lgalaxies} that a reservoir of gas transforms into stars according to
\beq
\rhosfr = \epsilon_{ff}\cdot\rhogas/t_{ff},
\eeq
where $\rhosfr$ is the star formation rate density, $\rhogas$ is the density of the gas in the reservoir, $\epsilon_{ff}$ is an efficiency per free-fall time, and $t_{ff}$ is the local dynamical time of the material \citep[see, e.g.,][for further discussion]{Krumholz_2014}. In our computation, we will dispense with the complications of modeling the distribution of gas densities in the object that determine $t_{ff}$ and instead assume a constant density of star-forming material. Under this assumption, we define $\alpha_{\rm SFR}\equiv \epsilon_{ff}/t_{ff}$ as an overall effective parameter governing star formation (which notably has units of inverse time). With this assumption, we get
\beq
\sfr = \alpha_{\rm SFR}\cdot\Mgas.
\eeq
For a stellar population of mass $\Mstar$ born at time $t_0,$ we write
$$F_{\rm surv}(t-t_0) + F_{\rm return,gas}(t-t_0) + F_{\rm return,Z}(t-t_0)=1,$$ where $F_{\rm surv}$ is the fraction of $\Mstar$ that survives until time $t,$ $F_{\rm return,gas}$ is the remaining fraction that is returned to the gas reservoir (e.g., via winds from massive stars), and $F_{\rm return,Z}$ is the remaining fraction returned into metals. Thus over the timescale $\Delta t,$ an amount $$M_{\rm return}=\Mstar\cdot\Delta t\cdot\dmqdt{F_{\rm return,gas}}$$ is returned to the gas reservoir over time as the stellar population evolves.

Given these assumptions, we can integrate a simple set of ODEs to compute $\Mstar(t),$ the total stellar mass that ever formed in the galaxy, and separately, $\Mstar^{\rm surv}(t),$ total stellar mass that survives until time $t.$ The gas consumption function that emerges from this simple model, $\fcons(t),$ is just the negative of the time derivative of the total mass in stars ever formed from non-recycled matter. Assuming a Chabrier IMF \citep{chabrier_2003_imf}, we approximate $F_{\rm return,Z}$ as a step function that goes from zero to $7.66\%$ after $4$ million years. This number is the fraction of mass in stars between 8 and 20 solar masses less the fraction in stellar remnants \citep[computed according to the fitting function presented in][]{hearin_etal22_dsps}. The results of our computation are shown in Figure~\ref{fig:sam_dep}; solid curves show results for a few different values of $\alpha_{\rm SFR};$ dotted curves show the behavior of the functional form for $\fcons(t\vert\taucons)$ assumed in the \dstar model; curves corresponding to different values of $\alpha_{\rm SFR}$ and $\taucons$ are color coded as indicated the legend.

We can see from Fig.~\ref{fig:sam_dep} the for each value of  $\alpha_{\rm SFR},$ the \dstar parameter $\taucons=2.75/\alpha_{\rm SFR}$ corresponds to a function $\fcons(t\vert\taucons)$ that roughly approximates the SAM-like model. This indicates that the one-parameter family of functions for $\fcons(t)$ assumed by our model broadly captures the behavior of the gas consumption function that emerges from elementary SAMs. Since the accuracy of our SFH fits degrades in low-mass halos, a potential avenue for further improvement of Diffstar is to use SAMs targeting dwarf galaxy growth \citep[e.g.,][]{kravtsov_manwadkar_grumpy21} to devise modifications to our assumed form of the consumption function. We are currently using the Galacticus SAM \citep{benson_galacticus_2012} to study how different features of SAMs manifest in different behaviors for $\fcons(t),$ as this will inform further refinements of the treatment of gas consumption in our model, but we leave this as a task for future work.


\bsp	
\label{lastpage}
\end{document}